\documentclass[hyper,letterpaper,12pt]{article}
\usepackage{mathrsfs}

\usepackage{amsfonts}
\usepackage{CJK}
\usepackage{a4wide}
\usepackage{amsmath}
\RequirePackage[dvips]{graphicx}
\usepackage[colorlinks=true,
            citecolor=blue,
            urlcolor=blue,
            filecolor=black,
            linktocpage=true,
            linkcolor=blue,
             ]{hyperref}

\def\nn{{\nonumber}}
\def\beq{\begin{equation}}
\def\eeq{\end{equation}}
\def\p{\partial}
\def\l{{\lambda}}
\def\L{{\Lambda}}
\def\b{{\beta}}
\def\a{{\alpha}}
\def\g{{ \gamma}}
\def\d{{\delta}}
\def\th{{\theta}}
\def\e{{\epsilon}}

\def\s{{ \sigma}}
\def\sc{$\Sigma_c$}

\def\t{\tau}
\def\m{\mu}
\def\n{\nu}
\def\hf{\frac{1}{2}}
\def\na{\nabla}

\def\pa{\partial}

\def\txi{\tilde {\xi}}

\def\tch{\tilde {\chi}}
\def\tp{\tilde {\psi}}
\def\tz{\tilde {\zeta}}
\def\tga{\tilde {\gamma}}
\def\tni{\tilde {n}}

\def\hG{\hat G}

\def\hn{\hat D}

\def \P{\mathcal {P}}
\def\mF{\mathcal {F}}

\def\mK{\mathcal {K}}
\def\mA{\mathcal {A}}
\def\mT{\mathcal {T}}
\def\mJ{\mathcal {J}}
\def\mG{\mathcal {G}}
\def\mR{\mathcal {R}}
\def\mN{\mathcal {N}}

\def\tmGr{\tilde{\mathcal {G}}^{\rho}}
\def\Pc{\tilde{\mathcal {P}}}
\def\hP{\tilde {\mathcal {P}}}
\def\hT{\tilde {\mathcal {T}}}
\def\hF{\tilde {\mathcal {F}}}
\def\hH{\tilde {\mathcal {H}}}
\def\hK{\tilde {\mathcal {K}}}
\def\hA{\tilde {\mathcal {A}}}
\def\hJ{\tilde {\mathcal {J}}}
\def\tmR{\tilde {\mathcal {R}}}
\def\hQ{\tilde {\mathcal {Q}}}
\def\hN{\tilde {\mathcal {N}}}
\def\V{V}

\def\H{\tilde{H}}
\def\E{\tilde{E}}
\def\W{\tilde{W}}
\def\R{\tilde{R}}
\def\tg{\tilde{g}}
\def\T{\tilde{T}}
\def\z{\tilde{\zeta}}

\def\A{\tilde{A}}
\def\K{\tilde{K}}
\def\F{\tilde{F}}

\def\f{\tilde{f}}
\def\ts{\tilde{s}}
\def\x{\tilde{x}}
\def\kr{\tilde{r}}
\def\u{\tilde{u}}
\def\C{\tilde{C}}
\def\N{\tilde{N}}
\def\J{\tilde{J}}

\def\tP{\tilde{P}}
\def\tL{\tilde{\Lambda}}

\def\si{\tilde{\sigma}}
\def\pc{\tilde{\partial}}
\def\tc{\tilde{\tau}}
\def\tm{\tilde{\mu}}
\def\tn{\tilde{\nu}}

\def\tO{\tilde{\Omega}}
\def\tr{\tilde{\rho}}
\def\tp{\tilde{p}}

\def\te{\tilde{\eta}}
\def\tz{\tilde{\zeta}}
\def\tnq{\tilde{n}}
\def\tth{\tilde{\theta}}
\def\tPh{\tilde{\Phi}}
\def\cc{\tilde{c}}

\def\cl{{\cal L}}

\def\BY{Brown-York}
\def\ri{\rightarrow}

\def\qq{\qquad}

\def\o{w}
\def\to{\tilde{w}}
\def\tmGo{\tilde{\mathcal {G}}^{w}}

\renewcommand{\(}{\left(}
\renewcommand{\)}{\right)}
\renewcommand{\[}{\left[}
\renewcommand{\]}{\right]}

\begin{document}
\title{\bf \Large Holographic Forced  Fluid Dynamics\\ 
in  Non-relativistic Limit}
\author{\normalsize ~Rong-Gen Cai~\footnote{E-mail: cairg@itp.ac.cn},
~Li Li~\footnote{E-mail: liliphy@itp.ac.cn}, ~Zhang-Yu Nie
\footnote{E-mail: niezy@itp.ac.cn}~ and Yun-Long
Zhang~\footnote{E-mail: zhangyl@itp.ac.cn}
\\
\\
\small State Key Laboratory of Theoretical Physics,\\
\small Institute of Theoretical Physics, Chinese Academy of Sciences,\\
\small Beijing 100190, People's Republic of China}
\date{\small August 3, 2012}
\maketitle

\begin{abstract}
\normalsize
 We study the thermodynamics and
non-relativistic hydrodynamics of the holographic fluid on a finite
cutoff surface in the Gauss-Bonnet gravity. It is shown that the
isentropic flow of the fluid is equivalent to a radial component of
gravitational field equations. We use the non-relativistic fluid
expansion method to study the Einstein-Maxwell-dilaton system with a
negative cosmological constant, and obtain the holographic
incompressible forced Navier-Stokes equations of the dual fluid at
AdS boundary and at a finite cutoff surface, respectively. The
concrete forms of external forces are given.
\end{abstract}

\tableofcontents

\section{Introduction}
\label{sect:intro}

Using the fluid/gravity
correspondence~\cite{Bhattacharyya:2008jc,Rangamani:2009xk}, it was
suggested that the problem of Navier-Stokes (NS) turbulence might be
mapped to a problem in general relativity
\cite{Bhattacharyya:2008ji,Bhattacharyya:2008kq}, with different
scales appearing in turbulent phenomena corresponding to different
radii in the dual geometry~\cite{Bredberg:2010ky}. Thus, with the
holographic Wilsonian  renormalization group (RG) approach
\cite{Iqbal:2008by,Heemskerk:2010hk,Faulkner:2010jy}, it is also
interesting to study the holographic hydrodynamics at a finite
cutoff surface
directly~\cite{Bredberg:2010ky,Iqbal:2008by,Matsuo:2011fk}. In this
study the timelike cutoff surface plays an important role so that
the asymptotical behavior of spacetime geometry becomes irrelevant.
This provides a potential approach to study holography in
asymptotically flat spacetimes.

It was proposed in \cite{Bredberg:2011jq} that there is a
mathematically precise relationship between the unforced
incompressible  NS equations in $p+1$ dimensions and vacuum Einstein
equations in $p+2$ dimensions and their solutions.  The dual
geometry has an intrinsically flat timelike cutoff surface whose
extrinsic curvature is identified as the stress energy tensor of the
dual fluid. This relationship has been further developed in the
literature such as
\cite{Compere:2011dx,Cai:2011xv,Lysov:2011xx,Bredberg:2011xw,Niu:2011gu,
Anninos:2011zn,Huang:2011he,Huang:2011kj}.
In \cite{Bredberg:2011jq}, a gravitational shock wave was introduced
to stir the fluid and then left to evolve according to the unforced
NS equations. In order to study the stationary NS turbulence, it is
better to introduce the external random source fields
\cite{Eling:2009pb,Eling:2010vr}.
For example, a dilaton field was added to the bulk gravity in
\cite{Bhattacharyya:2008ji}. In this case,  a perturbed gravity
solution with a slowly varying dilaton leads to a slowly varying
force term in NS equations.  Another example is
\cite{Bhattacharyya:2008kq}, where the force terms come from the
slow variation of the boundary background  in the holographic
context. It turns out that a simple forced steady state shear
solution to the forced NS equations becomes unstable and may
translate into turbulence at high enough Reynolds number.

In the fluid/gravity correspondence, the derivative expansion method
\cite{Bhattacharyya:2008jc} leads to the equations of motion of
relativistic fluid, and the equations reduce to the NS equations in
the non-relativistic limit~\cite{Bhattacharyya:2008kq,Eling:2009sj}.
Actually, the NS equations can also be obtained via taking the
non-relativistic expansion method directly
\cite{Bredberg:2011jq,Compere:2011dx}. This method can be used not
only for dual fluid at the AdS boundary, but also at  a finite
cutoff surface with flat induced
metric~\cite{Cai:2011xv}. In the latter case, the
bulk geometry is not required to be asymptotically AdS.
While in the
asymptotically AdS case, the perturbed gravity solutions with
Dirichlet condition at the cutoff surface can be mapped to the
perturbed gravity solutions without the cutoff surface
\cite{Kuperstein:2011fn,Brattan:2011my}. To study the holographic
fluid with external forces in non-relativistic limit, instead of
using the induced metric perturbations method proposed in
\cite{Bhattacharyya:2008kq,Brattan:2011my}, we keep the induced
metric flat and add external matter fields in the bulk as the source
terms to the dual fluids.

It was shown in \cite{Bredberg:2010ky} that the radial component of
Einstein equations is equivalent to the isentropy equation of the
dual fluid. Using a more generic static metric in some sense, we
show that this statement is also true in the Gauss-Bonnet gravity.
Using the non-relativistic expansion method, we  give the procedure
to obtain the perturbed solutions of the bulk gravity with a finite
cutoff surface up to the second order of non-relativistic expansion
parameter, and corresponding NS equations of the dual fluid at the
cutoff surface. It turns out that the shear viscosity over the
entropy density of the fluid dual to the Gauss-Bonnet gravity does
not run with the cutoff surface. This part acts as the service to
introduce the non-relativistic expansion method. In this paper we
mainly focus on an Einstein-Maxwell-dilaton system with a negative
cosmological constant and pay attention to the external force terms
in the dual non-relativistic fluid. The external force terms come
from the Maxwell field and dilaton field in the system.  Note that
the forced fluids at the AdS boundary have been discussed in
 the Einstein-dilaton system \cite{Bhattacharyya:2008ji} and
 Einstein-Maxwell system~\cite{Erdmenger:2008rm,Banerjee:2008th,Hur:2008tq},
 respectively. The perturbed solutions have been obtained to the second order
of the derivative expansion. We consider the
Einstein-Maxwell-dilaton system and obtain the perturbed solutions
up to the second order of the non-relativistic expansion
with/without the cutoff surface. Associated forced NS equations are
also derived in both cases. The concrete expressions of external
forces of the dual fluid, which come from the Maxwell field and dilaton field,
are given. The results show that the
Reynolds number of the dual fluid becomes larger and larger
 when the cutoff surface approaches the horizon of the background black branes.

This paper is organized as follows. In Section \ref{sect:review}, we
 start with a generic static black brane metric and obtain the
 perturbed solutions with a finite cutoff surface in the Gauss-Bonnet gravity by using the
 non-relativistic expansion method. This section acts as to fix the
 notations in this paper and to introduce the non-relativistic expansion
 method.  In Section \ref{sect:boundary}, we consider the Einstein-Maxwell-dilaton system
 with a negative cosmological constant and discuss the dynamics of
 dual fluid on the AdS boundary in non-relativistic limit. We
 generalize the discussions to the case with a finite cutoff
 surface in Section \ref{sect:cutoff}. The conclusions are given in Section \ref{sect:co}.

\section{Holographic fluid at a finite cutoff surface}
\label{sect:review} This section is a generalization of discussions
in \cite{Cai:2011xv} on the thermodynamics and hydrodynamics of dual
fluid at a finite cutoff surface.  Slight differently, we start with
a more general static metric and
 work in the intrinsic coordinates on the cutoff surface directly.

%
%
%
%
\subsection{Thermodynamics of the dual fluid at the cutoff surface}
To study a fluid in a $(p+1)$-dimensional flat spacetime, we
consider the generic $(p+2)$-dimensional static background
\begin{equation}\label{gm1}
ds_{p+2}^{2}=-g_{tt}(r)d t^{2}+g_{rr}(r)d
r^{2}+g_{xx}(r)\delta_{ij}dx^{i}dx^{j},
\qq \{i,j,...\}={1,2,...,p},
\end{equation}
where the metric components are  functions of radial coordinate $r$
only. We assume the  metric has a well-defined event horizon at
$r=r_h$, where $g_{tt}(r)$ has a first order zero $g_{tt}(r_h)=0$,
 and $g_{rr}(r)$ has a first order pole $g_{rr}^{-1}(r_h)=0$~\cite{Iqbal:2008by}.
For example, the ingoing-Rindler form of flat spacetime and the
black $p$-brane solutions in asymptotically AdS spacetime have the
form~\cite{Bredberg:2010ky}. Using the Eddington-Finkelstein
coordinate $\t$ defined by $d\t = dt +  \sqrt{g_{rr}(r)/g_{tt}(r)}
{dr }$ \cite{Iqbal:2008by}, we can rewrite the metric (\ref{gm1}) as
\begin{equation} \label{EFC}
ds_{p+2}^{2}=2\sqrt{g_{tt}(r)g_{rr} (r)} d\t dr-g_{tt}(r)d \t^{2}+g_{xx}(r)\delta_{ij}dx^{i}dx^{j},
\end{equation}
which has the translational invariance in $\t$ and $x^i$ directions.
We can always introduce a finite cutoff surface \sc ~at $r=r_c$
outside the horizon with the intrinsic coordinates, $\x^a \sim
(\tc,\x^i)$, as
\begin{equation}\label{2.3}
\x^0\equiv\tc=\sqrt{g_{tt}(r_c)}~\t,\qq \x^i=\sqrt{g_{xx}(r_c)}~x^i,
\qq \{a,b...\}={0,1,2,...,p}.
\end{equation}
Then the associated bulk metric (\ref{EFC}) becomes
\begin{align}\label{bulk}
ds_{p+2}^{2}
&=g_{rr}(r) d r^2 +{\g}_{ab}(r)
\[d\x^a+\mN(r) \d^a_{\tc} d r\]
\[d\x^b+\mN(r) \d^b_{\tc} d r\],
\end{align}
where ${\g}_{ab}(r)$ and $\mN(r)$ are given by
\begin{align} \label{hNr}
&{\g}_{ab}(r)d\x^a d\x^b=
-\frac{g_{tt}(r)}{g_{tt}(r_c)}d\tc^2+\frac{g_{xx}(r)}{g_{xx}(r_c)}d\x_i d\x^i,
\qq \mN(r)=-\frac{\sqrt{g_{tt}(r_c)g_{rr}(r)}}{\sqrt{g_{tt}(r)}}.
\end{align}
And the induced metric with intrinsic coordinates on $\Sigma_c$ is
simply given as
\begin{equation} \label{flat}
d\ts_{p+1}^2={\g}_{ab}(r_c)d\x^ad\x^b=\tilde{\eta}_{ab}d\x^ad\x^b=-d\tc^2+\d_{ij}d\x^id\x^j.
\end{equation}

The Brown-York stress energy tensor $\hT^{BY}_{ab}$ evaluated on the
cutoff hypersurface $\Sigma_c$ is proposed as the stress energy
tensor of the dual fluid~\cite{Bredberg:2011jq}. It has a close
relation with the extrinsic curvature tensor
$\K_{ab}=\frac{1}{2}\cl_{\N}{\g}_{ab}(r)|_{r=r_c}$
 of $\Sigma_c$,
where $\cl_{\N}$ is the Lie derivative along the unit normal $\N^A$
of the hypersurface. To study the thermodynamic properties of the
dual fluid at the cutoff surface, we begin with the re-scaled metric
(\ref{bulk}) with a Killing horizon located at $r=r_h$. The local
temperature $\T_{{_{0}}}(r_c)$ on \sc, which is identified as the
temperature of dual fluid~\cite{Bredberg:2010ky}, meets the Tolman
relation with Hawking temperature $T_H$ of the black brane metric
(\ref{EFC})
\begin{equation}
\begin{split}\label{TH}
\T_{{_{0}}}(r_c)=\frac{T_H}{\sqrt{g_{tt}(r_c)}},\qq
T_H\equiv\lim_{r\rightarrow
r_h}\frac{g'_{tt}(r)}{4\pi\sqrt{g_{tt}(r)g_{rr}(r)}}
\end{split}
\end{equation}
To discuss the local entropy density of the dual fluid,
we consider a quotient of the general geometry (\ref{EFC}) under shift of $x^i$\cite{Bredberg:2010ky},
$x^i\sim x^i+\ell_0 n^i$, with a characteristic length $\ell_0$  and $n^i\in Z$.
Equivalently, using metric (\ref{bulk}), the spatial $R^p$ on \sc~turns out to be a $p$-tours $T^p$ with $r_c$-dependent volume
\begin{equation}
V_p(r_c)=g_{xx}^{p/2}(r_c)V_0,\qq V_0=\ell^{p}_0,\qq \x^i\sim \x^i+\ell_0\sqrt{g_{xx}(r_c)} n^i.
\end{equation}
As the total Bekenstein-Hawking entropy $S$ is fixed, we can
identify the dual  fluid's entropy density as
\begin{equation}
\begin{split}\label{s0}
s_{_{0}}(r_c)=\frac{S}{V_p(r_c)}=\frac{1}{4G_{p+2}}\frac{g_{xx}^{p/2}(r_h)}{g_{xx}^{p/2}(r_c)},
\qq S=\frac{V_p(r_h)}{4G_{p+2}},
\end{split}
\end{equation}
where $S$ is the Bekenstein-Hawking entropy of the black brane.

As a calculation example, we consider the Einstein-Hilbert action
with a non-positive cosmological constant $\Lambda\leq0$, as well as
the Gauss-Bonnet term
$\mathcal{L}_\mathrm{GB}=R^2-4R_{AB}R^{AB}+R_{ABCD}R^{ABCD}$, and
the contribution from matters $\mathcal {L}_\mathrm{M}$,
\begin{eqnarray}
\label{action} I=\frac{1}{16 \pi G_{p+2}}\int
~d^{p+2}x\sqrt{-g} \left(R-2 \Lambda+\alpha\mathcal {L}_{\mathrm{GB}}+\mathcal {L}_\mathrm{M}\right),
\end{eqnarray}
where $\alpha$ is the Gauss-Bonnet coefficient with the same
dimension as square of length. The Gauss-Bonnet term is a
topological invariant when $p=2$, and will modify Einstein
gravitational field equations when $p\geq3$~\cite{Deser,Cai:2001dz}. We
will generally demand $p\geq2$ and set $16 \pi G_{p+2}=1$ throughout
this paper~\footnote{Most of the results in this paper have been
checked by Mathematica up to $p=5$.}. Varying the action
(\ref{action}) with respect to the induced metric $\g^{ab}$ with
appropriate surface terms~\cite{Myers:1987yn,Davis:2002gn,Astefanesei:2008wz,Brihaye:2008xu}, we can
get the Brown-York stress energy tensor $\hT^{BY}_{ab}$ evaluated on
the cutoff hypersurface $\Sigma_c$ of the Gauss-Bonnet gravity.
Because of the flatness of the cutoff surface, its final form is
given by \cite{Cai:2011xv}
\begin{equation}\label{BYdef}
{\hT^{BY}_{ab}=2 \left[\K\te_{ab}-\K_{ab}
- 2\alpha \(3\J_{ab}-\J\te_{ab}\)+\C\te_{ab} \right],}
\end{equation}
%
with
\begin{equation}
\J_{ab}=\frac{1}{3}%
(2\K\K_{ac}\K_{~b}^{c}+\K_{cd}\K^{cd}\K_{ab}-2\K_{ac}\K^{cd}\K_{db}-\K^{2}\K_{ab})~,
\label{Jab}
\end{equation}
where $\K$ is the trace of extrinsic curvature tensor $\K_{ab}$ of
$\Sigma_c$, and $\C$ is an ambiguous constant. Substituting the
metric (\ref{bulk}) into the \BY~stress tensor (\ref{BYdef}), we can
get the stress tensor of the holographic fluid dual to the
Gauss-Bonnet gravity as
\begin{equation}
\begin{split}\label{static}
&\hT^{BY}_{ab}d\x^ad\x^b=\[\tr_{_{G}}(r_c)-2\C\]d\tc^2+\[\tp_{_{G}}(r_c)+2\C\]\delta_{ij}d\x^id\x^i,
\end{split}
\end{equation}
with constant pressure $\tp_{_{G}}(r_c)+2\C$ and energy density
$\tr_{_{G}}(r_c)-2\C$, where
\begin{equation}
\begin{split}\label{BYC1}
&\tr_{_{G}}(r_c)=\tmGr(r_c)\tr_{_{E}}(r_c),
\qq\qq~~\to_{_{G}}(r_c)\equiv\tr_{_{G}}(r_c)+\tp_{_{G}}(r_c)=\tmGo(r_c)\to_{_{E}}(r_c),\\
&\tr_{_{E}}(r_c)=-\frac{p}{\sqrt{g_{rr}(r_c)}}\frac{g'_{xx}(r_c)}{g_{xx}(r_c)},
\qq\to_{_{E}}(r_c)\equiv
\frac{1}{\sqrt{g_{rr}(r_c)}}\(\frac{g'_{tt}(r_c)}{g_{tt}(r_c)}-\frac{g'_{xx}(r_c)}{g_{xx}(r_c)}\).
\end{split}
\end{equation}
The sum $\to_{_{G}}(r_c)$ is independent of the constant $\C$. 
The Gauss-Bonnet term corrections, compared to Einstein gravity with
$\l_{G}=(p-1)(p-2)\a$,  are given by
\begin{align}
&\tmGr(r_c)=1-\frac{\l_{G}}{6{g_{rr}(r_c)}}\(\frac{g'_{xx}(r_c)}{g_{xx}(r_c)}\)^2,\qq
\tmGo(r_c)=1-\frac{\l_{G}}{2{g_{rr}(r_c)}}\(\frac{g'_{xx}(r_c)}{g_{xx}(r_c)}\)^2.
\end{align}

Using the above results, in general we  have the thermodynamic
relation
\begin{equation}
\begin{split}
&\T_{_{0}}(r_c)s_{_{0}}(r_c)=\to_{_{G}}(r_c)-\tO_{_{0}}(r_c),
\qq\tO_{_{0}}(r_c)=\sum_m\m^{(q)}_m (r_c) \tnq_{^{(q)}}^m(r_c),
\end{split}
\end{equation}
where $m=1,2,...$, and every $\tnq_{^{(q)}}^m(r_c)$ corresponds to
different kinds of charge density with corresponding  conjugate
chemical potential $\m^{(q)}_m (r_c)$. Note that in the Gauss-Bonnet
theory, the area formula of black hole entropy no longer holds; but
for the black brane solution with Ricci flat horizon considered in
the present paper, the area formula for the black brane entropy
still works~\cite{Jacobson:1993xs,Cai:2001dz,Brigante:2007nu}. Thus
the total entropy is given by \footnote{In the fluid/gravity
correspondence, the corresponding Wald entropy current of
curvature-squared gravity theories is studied in
\cite{Chapman:2012my}.}
\begin{eqnarray}\label{entropy}
S=s_{_{0}}(r_c)V_p(r_c)=\frac{V_p(r_c)}{\T_{_{0}}(r_c)}\[\to_{_{G}}(r_c)-\tO_{_{0}}(r_c)\],
\qq\pa_{ r_c} S=0.
\end{eqnarray}
This isentropic equation $\pa_{ r_c} S=0$ can be considered as
either an adiabatic thermodynamic process of the dual fluid or a
holographic renormalization group flow~\cite{Bredberg:2010ky}.
Varying the action (\ref{action}) with respect to metric $\tg^{AB}$,
we can obtain the equations of motion for the Gauss-Bonnet gravity
in the re-scaled (p+2)-dimensional bulk coordinates
\begin{equation}
\begin{split}\label{WAB}
&\W_{AB}(r)\equiv\E_{AB}(r)+\alpha \H_{AB}(r)-\frac{1}{2}~
\T_{AB}(r)=0, \qq \{A,B...\}=\{r,\tc,\x^i\}
\end{split}
\end{equation}
with the definitions
\begin{eqnarray}
\begin{aligned}
\label{GBeqs}
&\E_{AB}(r)\equiv\R_{AB}-\frac{1}{2}\R ~\tg_{AB}+\L\tg_{AB},
\qq\T_{AB}(r)\equiv -2 \frac{\delta \mathcal{L}_\mathrm{M}}{\delta \tg^{AB}} +\tg_{AB} \mathcal{L}_\mathrm{M},
\\
&\H_{AB}(r)\equiv2(\R_{A CDE}\R_{B }^{\phantom{B}CDE}-2\R_{A C B D}\R^{CD}-2\R_{A
C}\R_{\phantom{C}B }^{C }+\R\R_{AB})
-\frac{1}{2}\tg_{AB}\mathcal {L}_\mathrm{GB},
\end{aligned}
\end{eqnarray}
where $\T_{AB}(r)$ is the stress energy tensor of bulk matters. If
we assume the metric (\ref{bulk}) solves Eqs. (\ref{WAB}), the
nonzero components of the stress energy tensor  are
$\T_{rr}(r),\T_{\tc\tc}(r),\T_{\x\x}(r)\equiv\T_{\x^i\x^i}(r)$,
 and $\T_{\tc r}(r)=\T_{r\tc}(r)=\mN(r)\T_{\tc\tc}(r)$, respectively.
Following the procedure in \cite{Bredberg:2010ky} and using metric
(\ref{bulk}), we can introduce an arbitrary null vector $\z$
tangent to the hypersurface $r=r_c$ with time component $\p_{\tc}$,
such as
\begin{equation}\label{TMN}
\z^A\p_A=\p_{\tc}-\p_{\x^1}
\qq\tg_{AB}\z^A\z^B=0,
\qq \z^A\N_A=0,
\end{equation}
with $\tilde N$ the unit normal of the cutoff surface. It could be
checked that
\begin{equation}
\begin{split}
&\pa_{ r_c}\[\frac{V_p(r_c)}{\T_{_{0}}(r_c)}\to_{_{G}}(r_c)\]
=\frac{V_p(r_c) }{T_H}\sqrt{g_{rr}(r_c)g_{tt}(r_c)}~2~\z^A\z^B\[\E_{AB}(r_c)+\a\H_{AB}(r_c)\].
\end{split}
\end{equation}
 When the $\tO_{_{0}}(r_c)$ term satisfy the following relation
\footnote{For an example see Section \ref{sect:cutoff} of this paper or
\cite{Bredberg:2010ky}. For the vacuum case without the matter
source in (\ref{action}), this relation always holds.}
\begin{equation}
\begin{split}
&\pa_{ r_c}\[\frac{V_p(r_c)}{\T_{_{0}}(r_c)}\tO_{_{0}}(r_c)\]=
\frac{V_p(r_c) }{T_H}\sqrt{g_{rr}(r_c)g_{tt}(r_c)}~\z^A\z^B\T_{AB}(r_c).
\end{split}
\end{equation}
 Using
(\ref{entropy}),(\ref{WAB}) and the above two equations, we have
\begin{equation}
\pa_{ r_c} S=\frac{V_p(r_c) }{T_H}\sqrt{g_{rr}(r_c)g_{tt}(r_c)}~2~\z^A\z^B\W_{AB}(r_c).
\end{equation}
As $\sqrt{g_{rr}(r_c)g_{tt}(r_c)}$, ~$V_p(r_c)$ and $T_H$ have
neither zero nor pole even when $r_c=r_h$, we arrive at
\begin{equation}
\pa_{ r_c} S=0~\Longleftrightarrow ~\z^A\z^B \W_{AB}(r_c)=0,
\end{equation}
which implies the equivalence between the isentropy of the RG flow
and a radial gravitational field equation of the Gauss-Bonnet
gravity. Thus we have generalized the result in
\cite{Bredberg:2010ky} to the case of Gauss-Bonnet gravity.

\subsection{Incompressible Navier-Stokes equations from gravity}

Keeping the intrinsic metric (\ref{flat}) of  \sc\  flat, we can
take two linear diffeomorphisms based on metric (\ref{bulk}): a
transformation of the radial $r$ and a Lorentz boost in $(\tc,
~\x^i)$ coordinates. The first is the transformation of $r$ and the
associated re-scaling of $(\tc, ~\x^i)$ as
\begin{equation}
 r \rightarrow k(r),~~~~\tc\rightarrow \tc \sqrt{\frac{g_{tt}(r_c)}{g_{tt}[k(r_c)]}}
,~~~~~\x^i\rightarrow \x^i\sqrt{\frac{g_{xx}(r_c)}{g_{xx}[k(r_c)]}},
 \end{equation}
where $k(r)$ is a linear function of $r$, whose  form would  be
chosen via the global symmetry of geometry (\ref{bulk}). The metric
(\ref{bulk}) is then transformed into
\begin{equation} \label{scale0}
ds_{p+2}^{2}\rightarrow d\hat{s}_{p+2}^{2}= 2\frac{\sqrt{g_{tt}(\kr)g_{rr}(\kr)}}{\sqrt{g_{tt}(\kr_c)}}
d\tc d \kr-\frac{g_{tt}(\kr)}{g_{tt}(\kr_c)}d
\tc^{2}+\frac{g_{xx}(\kr)}{g_{xx}(\kr_c)}\delta_{ij}dx^{i}dx^{j}.
\end{equation}
where we have introduced the re-scaled coordinate  $\kr\equiv k(r)$
and the notation $\kr_c\equiv k(r_c)$. We will work in the
$(r,\x^a)$ coordinates directly in this paper.

The other diffeomorphism is the Lorentz boost with a constant boost
parameter $\beta_i$ in the $\x^a=(\tc, ~\x^i)$ coordinates
\begin{equation}\label{bt}
\tc \rightarrow -\u_{a}\x^a,
\qq\x^i \rightarrow \tni_{a}^i\x^a,
\qq \tni_{a}^i=\(-\g\beta^i,~\delta^i_j+(\g-1)\frac{\beta^i\beta_j}{\beta^2}\),
\end{equation}
where we also have defined the $(p+1)$-velocity
\begin{equation}\label{t1}
\u_{a}=\g_{_{\beta}}(-1, ~\beta_i),\qq
\g_{_{\beta}}=\(1-\beta^2\)^{-\frac{1}{2}},
\qq\beta^2=\beta_i\beta^i=\d_{ij}\beta^i\beta^j.
\end{equation}
As these parameters are all constants,
the associated transformations 
can be expressed as
\begin{equation}
 d\tc \rightarrow -\u_{a} d\x^a,~~~
 d \x^i  \rightarrow \tni^i_{a} d\x^a, ~~~
 \delta_{ij}d \x^id \x^j  \rightarrow \delta_{ij}\tni^i_{a}\tni^j_{a} d\x^a d\x^b=\tilde{\P}_{ab}d\x^a d\x^b,
\end{equation}
with the projection operator $
\tilde{\P}_{ab}\equiv\te_{ab}+\u_{a}\u_{b}=\delta_{ij}n_{a}^in_{b}^j
$. Then the metric (\ref{scale0}) becomes
\begin{equation} \label{boost}
\begin{split}
d\hat{s}_{p+2}^{2}\rightarrow d\ts_{p+2}^{2}&
=g_{rr}(\kr) d \kr^2 +{\tga}_{ab}(\kr)
\[d\x^a+\hN(\kr)\u^a d \kr\]
\[d\x^b+\hN(\kr)\u^b d \kr\],
\end{split}
\end{equation}
where we have used the ADM-like decomposition
at the constant $\kr$ surface,
\begin{equation}
\begin{split}
\tilde{\g}_{ab}(\kr)&=
-\frac{g_{tt}(\kr)}{g_{tt}(\kr_c)}\u_{a}\u_{b}+\frac{g_{xx}(\kr)}{g_{xx}(\kr_c)}
\tilde{\P}_{ab},
\qq \hN(\kr)=-\frac{\sqrt{g_{tt}(\kr_c)g_{rr}(\kr)}}{\sqrt{g_{tt}(\kr)}}.
\end{split}
\end{equation}
%
After the two diffeomorphisms, the metric (\ref{bulk})  and
 (\ref{boost}) still solve the same gravity field equations.
The cutoff surface $r=r_c$ is equivalent to  $\kr=\kr_c$, we will
firstly work with the new radial variable $\kr$ , and then transform
back to $r$ via $r=k^{-1}(\kr)$.

In general, the \BY~ stress energy tensor on the cutoff surface
$r=r_c$ corresponding to metric (\ref{boost}) will turn out to take
the form
\begin{equation}\label{ideal}
\hT^{BY}_{ab}(r_c)=\tr(r_c)\u_a\u_b+\tp(r_c)\hP_{ab},\qq
\to(r_c)=\tr(r_c)+\tp(r_c),
\end{equation}
which could be identified as the stress tensor of an ideal
relativistic fluid  $\hT^{(ideal)}_{ab}$ with $(p+1)$-velocity $\u_a$
in flat space-time. In the Gauss-Bonnet gravity case (\ref{BYdef}),
the cutoff-dependent energy density
$\tr(r_c)\equiv\tr_{_G}(\kr_c)-2\C$ and pressure
$\tp(r_c)\equiv\tp_{_G}(\kr_c)+2\C$. In general here $\C$ is an
unfixed constant, but to obtain a finite result when the cutoff
surface goes to the AdS boundary, $\C$ can be
fixed~\cite{Cai:2011xv}.

Let us pause to  recall the hydrodynamical description of
microscopic field dynamics in flat spacetime. It applies when the
correlation length of the fluid $l_{cor}$ is much smaller than the
characteristic scale $L$ of variations of the macroscopic fields
\cite{Fouxon:2008tb}. Via dimensional analysis, $l_{cor}\sim1/T_c$
and $1/L\sim\p_{\x^a}$, where $T_c$ is the characteristic
temperature of the fluid and $\p_{\x^a}$ would act as the
coordinate-dependent parameters, such as $T(\x)$ and $\u^a(\x)$
\footnote{We have used $(\x)$ to denote the function arguments
$(\x^a)$, and would use $\pc_{a}$ to denote $\p_{\x^a}$.}. One
introduces the dimensionless Knudsen number $Kn\equiv
l_{cor}/L\sim\frac{1}{T}\pc_{a}\sim\e\ll 1$ to expand the stress
energy tensor of relativistic fluid  in flat background
$\tilde{\eta}_{ab}$,
\begin{equation}
\begin{split}\label{Knudsen}
\hT_{ab}(\x)=\sum_{n=0}^{\infty}\hT^{(n)}_{ab}(\x),\qq
\hT^{(0)}_{ab}(\x)=\to(\x) \u_a(\x) \u_b(\x) +\tp(\x)
\tilde{\eta}_{ab},
\end{split}
\end{equation}
and $\hT^{(n)}_{ab}(\x)\sim(Kn)^n$. To take the non-relativistic
limit as in~\cite{Fouxon:2008tb}, we recover the speed of light $c$
in (\ref{Knudsen}) and introduce the normalized pressure $P(\x)$ as
\begin{equation}
\begin{split}\label{up}
&\u_a(\x)=\g_{_{\beta}}(\x)\(-1,{\b_i(\x)}/{c}\),~~~~
\g_{_{\beta}}(\x)={\(1-{\b^2(\x)}/{c^2}\)^{-\frac{1}{2}}},
~~~~\b^2(\x)=\b_i(\x)\b^i(\x),\\
&\u^a(\x)\pc_a=\g_{_{\beta}}(\x)\frac{\pc_{\t}}{c} +\g_{_{\beta}}(\x)\frac{\b_i(\x)}{c}\pc_i~,
~\frac{\pc_aT(\x)}{T(\x)}\sim\frac{\pc_a\to(\x)}{\to(\x)}\sim\frac{\pc_a\tp(\x)}{
\tp(\x)}\equiv\frac{\to(\x)}{\tp(\x)}\frac{\pc_a\tP(\x)}{c^2}.
\end{split}
\end{equation}
When $c\rightarrow\infty$, the energy-momentum conservation
equations of the ideal fluid, $\pc^a\hT^{(0)}_{ab}(\x)=0$, leads to
the non-relativistic incompressible Euler's equations,
\begin{equation}
\begin{split}\label{euler}
\pc_i \tP(\x)+\pc_{\t}\b_i(\x)+\b^j(\x)\pc_j\b_i(\x)=0,
 \qq \pc_i \b^i(\x)=0.
\end{split}
\end{equation}
Instead of the $c\rightarrow\infty$ limit, if we assume the small
velocity parameter to take the same limit as the Knudsen number,
i.e. $\b/c\sim Kn\sim\e$, it would  be  equivalent to set $c=1$ in
(\ref{up}) with the following scalings
\begin{equation}
\begin{split}\label{scale}
\pc_i\sim\e,\qq\pc_\t\sim\e^2,\qq  \b_i(\x)\sim\e, \qq \tP(\x)\sim\e^2.
\end{split}
\end{equation}
These non-relativistic scalings are named as the BMW limit
\cite{Bhattacharyya:2008kq,Brattan:2011my}. To obtain the forced NS
equations, the dissipative part of the stress energy tensor
(\ref{Knudsen}) is required
\begin{equation}\label{dissipative}
\hT_{ab}(\x)=\hT^{(0)}_{ab}(\x)+\hT^{(diss)}_{ab}(\x),\qq\hT^{(diss)}_{ab}(\x)=\hT^{(1)}_{ab}(\x)+\hT^{(2)}_{ab}(\x)+...~.
\end{equation}
For example, in the Landau frame $\u^a(\x)\hT^{diss}_{ab}(\x)=0$,
the first order dissipative components could be written as
\begin{equation}
\begin{split}\label{dis}
\hT^{(1)}_{ab}(\x)&=-2\tilde{\eta}(\x)\si_{ab}(\x)-\tilde{\zeta}(\x)\tilde{\theta}(\x)\Pc_{ab}(\x),
\qq\Pc_{ab}(\x)=\te_{ab}+\u_a(\x)\u_b(\x),\\
\si_{ab}(\x)&=\Pc_a^m(\x)\Pc_b^n(\x) \pc_{(m}\u_{n)}(\x)-\frac{\Pc_{ab}(\x)}{p}\tilde{\theta}(\x),
~~~\tilde{\theta}(\x)=\te^{ab}\pc_a\u_b(\x),
\end{split}
\end{equation}
where $\tilde{\eta}(\x)$ is the kinetic shear viscosity and
$\tilde{\zeta}(\x)$ is the bulk viscosity, the latter vanishes for
conformal fluids. Usually, they behave the same as local temperature
$\T(\x)$ in (\ref{up}). The first order dissipative hydrodynamics
satisfies the dynamical equations
$\pc^a[\hT^{(0)}_{ab}(\x)+\hT^{(1)}_{ab}(\x)]=\f_{b}(\x)$ if some
external source term appears. In the non-relativistic limit
(\ref{scale}), if we further assume
$\f_{i}(\x)\sim\e^3,\f_{\tc}(\x)\sim\e^4$, it would lead to the
non-relativistic forced incompressible NS equations at order $\e^3$,
\begin{equation}
\begin{split}\label{naivor}
\pc_i \tP(\x)+\pc_{\t}\b_i(\x)+\b^j(\x)\pc_j\b_i(\x)+\tn(\x)\pc^j\pc_j\b_i(\x)=\f^{[\to]}_i(\x),
 \qq \pc_i \b^i(\x)=0,
\end{split}
\end{equation}
where $\tn(\x)=\eta(\x)/\to(\x)$ is the dynamical shear viscosity
and $\f^{[\to]}_i(\x)=\f_i(\x)/\to(\x)$. The NS equations have the
scaling symmetry (\ref{scale}), as shown in
\cite{Bhattacharyya:2008kq,Bredberg:2011jq}.

Next we will derive the incompressible  NS equations from the
gravity side. To get the dissipative part of the dual fluids, we
need to perturb the geometry (\ref{boost}) by using either the
linear response method (eg.~\cite{Iqbal:2008by}) or the perturbative
expansion method developed in~\cite{Bhattacharyya:2008jc}, where a
perturbative procedure to solve Einstein's equations order by order
in the boundary derivative expansion was proposed. In the case with
a finite cutoff surface, we also can employ the procedure under the
non-relativistic limit~\cite{Compere:2011dx,Cai:2011xv}. Regarding
the transformation parameters in linear function $\kr=k(r)$ and the
$p$-velocity $\b^i$ in the boost transformation as  functions of the
hypersurface coordinates $(\tc,\x^i)$, we can solve the
gravitational equations order by order in the non-relativistic
perturbative expansion. Based on (\ref{scale}), these transformation
parameters are also regarded as small quantities with appropriate
scaling symmetry as
\begin{equation}
\begin{split}\label{nonrel}
\p_ r \sim \e^0,~~~\pc_{i} \sim \b_i(\x) \sim \e^1, ~~~\pc_{\t}\sim \d k(r,\x)\sim \tP(\x)\sim \e^2,
\end{split}
\end{equation}
where  $\d k(r)\equiv\kr-r\sim\e^2$ would provide the pressure
perturbation. Using the formal Taylor expansion such as
$g_{tt}(\kr)=g_{tt}(r)+g_{tt}'(r)\d k(r)$, and define $\tg_{\tc
r}(r)=\sqrt{g_{tt}(r)g_{rr}(r)/g_{tt}(r_c)} $, we can consider the
metric (\ref{boost}) with coordinate-dependent parameters as
$d\ts_{(0)}^2$, and expand it up to order $\e^2$,
\footnote{In this paper, we use the scripts $(0),(1),...$ to denote the order in the derivative expansion, 
and scripts $(\e),(\e^2)...$ to denote the order of the
non-relativistic expansion parameter $\e$.}
\begin{align}\label{dsexp}
d\ts_{(0)}^2 =&+ 2\tg_{\tc r} (r) d\tc dr
     + \[-\frac{g_{tt}(r)}{g_{tt}(r_c)}d\t^2 + \frac{g_{xx}(r)}{g_{xx}(r_c)}\delta_{ij}dx^{i}dx^{j}\]\nn \\
&  -2\tg_{\tc r} (r) \b_i(\x) d\x^i dr
     +2\[ \frac{g_{tt}(r)}{g_{tt}(r_c)}-\frac{g_{xx}(r)}{g_{xx}(r_c)}\]\b_i(\x)d\x^i d\tc\nn\\
& + \tg_{\tc r} (r) \b^2(\x) d\tc dr
    - \[ \frac{g_{tt}(r)}{g_{tt}(r_c)}-\frac{g_{xx}(r)}{g_{xx}(r_c)}\]\( \b^2(\x) d\tc^2+\b_i(\x) \b_j(\x) d\x^i d\x^j \)\nn\\
& +\tg_{\tc r} (r)\frac{g'_{rr}(r)}{g_{rr}(r)}\d k(r)d\tc dr +\[\frac{g'_{tt}(r)}{g_{tt}(r)}\d k(r)-\frac{g'_{tt}(r_c)}{g_{tt}(r_c)}\d k(r_c)\]\(\tg_{\tc r} (r)d\tc dr-\frac{g_{tt}(r)}{g_{tt}(r_c)}d\tc^2\)\nn\\
& +2\tg_{\tc r} (r)\[k'(r)-1\]d\tc dr  +\frac{g_{xx}(r)}{g_{xx}(r_c)}\[\frac{g'_{xx}(r)\d k(r)}{g_{xx}(r)}-\frac{g'_{xx}(r_c)}{g_{xx}(r_c)}\d k(r_c)\] ~d\x_id\x^i,\nn\\
& +O(\e^3),
\end{align}
where $\d k(r)$ and $\b_i(\x)$ and are all functions of $\x^a$ now,
we will omit this notation $(\x)$ henceforth. The coordinate-dependent metric (\ref{dsexp}) is no longer a diffeomorphism of
metric (\ref{boost}). Under the perturbative expansion, it turns out
that (\ref{dsexp}) only solves the same equations of motion of
gravity up to $\e^1$. To solve the equations of motion  up to
$\e^2$, some constraint equations and new correction terms to the
bulk metric are needed at this order. With the same way, the
equations of motion of the system can  be solved order by order in
the non-relativistic expansion parameter $\e$~\cite{Compere:2011dx}.
And the corresponding \BY~ stress energy tensor which is identified
as the dual fluid's stress energy tenor, can also be obtained at the
desired order. In this paper we solve the equations of motion up to
 order $\e^2$. It turns out that the non-dissipative part is still
given by (\ref{ideal}) in the non-relativistic limit. Up to $\e^2$,
we have
\begin{equation}
\begin{split}\label{ideal1}
\hT^{(0)}_{ab}d\x^a d\x^b&=\[\tr(r_c)+\to_{_0}(r_c)\b^2\] d\tc^2-2\to_0(r_c)\b_i d\x^id\tc\\
&+\[\tp(r_c)\d_{ij}+\to_{_0}(r_c)\b_i \b_j \]d\x^i d\x^j+O(\e^3),
\end{split}
\end{equation}
where $\to_{_0}(r_c)=\tr_{_0}(r_c)+\tp_{_0}(r_c)$, and
\begin{align}
\tr(r_c)&\equiv\tr_{_0}(\kr_c)\cong\tr_{_0}(r_c)+\tr'_{_0}(r_c)\d k(r_c),
~~~\tp(r_c)\equiv\tp_{_0}(\kr_c)\cong\tp_{_0}(r_c)+\tp'_{_0}(r_c)\d k(r_c).
\end{align}
The equations of motion of the bulk matters should be solved in the same procedure.
Once we get the solutions of both gravity and matters up to order $\e^2$,
the constraint equations of gravity at order $\e^3$ are just
the forced incompressible NS equations (\ref{naivor}) of the dual fluid.

As a calculation example to perturb the geometry (\ref{boost}), we again assume the metric 
solves the equations (\ref{WAB}) of Gauss-Bonnet gravity with a
non-positive cosmology constant $\tL$. The metric (\ref{dsexp})
leads to $\W^{(\e)}_{MN}=0$, and the constraint equation at order
$\e^2$ turns out to be
\begin{equation}\label{ein1}
  \qq 2\N^C\W^{(\e^2)}_{C\t}=-\pc^a\T^{(0)}_{a\t}=\to_{_0}(r_c)\pc_i\b^i=0,
\end{equation}
where $\to_{_0}(r_c)=\to_{_G}(r_c)$ is nonzero outside the horizon.
Thus it leads to the incompressible condition $\pc_i\b^i=0$  of the
dual fluid at the cutoff surface.

To solve gravitational field equations at order $\e^2$, we need to
add corrections to the metric (\ref{dsexp}). It was shown in
\cite{Bhattacharyya:2008jc} that due to the spatial $SO(p)$ rotation
symmetry of the black brane background, one has the decoupled
equations of $SO(p)$ scalar, vector and traceless tensor
perturbations. For example, if we turn off the tensor perturbations
of the matter stress energy tensor $\hT_{AB}(r)$, the tensor
corrections to the metric, at order $\e^2$, are
\begin{align}\label{gij1}
d\ts^{2}_{(1)}=2\hF(r)\frac{g_{xx}(r)\sqrt{g_{tt}(r_c)}}{g_{xx}(r_c)}\si_{ij}~d\x^i d\x^j+...,
\qq \si_{ij}=\pc_{(i}\b_{j)}-\frac{1}{p}\d_{ij}\pc_k\b^k.
\end{align}
Then the traceless tensor parts of the gravitational field equations
$\W^{(\e^2)}_{ij}=0$ lead to a second-order ordinary differential
equation
\begin{align}\label{hFr}
\[\(1+\frac{\sqrt{g_{tt}(r)}}{\sqrt{g_{rr}(r)}}\hF'(r)\)\mG_g(r)\]'=0
\Rightarrow
\hF(r)=\int_r^{r_c}\frac{\sqrt{g_{rr}(y)}}{\sqrt{g_{tt}(y)}}\[1-\frac{\mG_{g}(r_h)}{\mG_{g}(y)}\]dy,
\end{align}
where two integration constants have been fixed and
\begin{equation}\label{eta}
\begin{split}
\mG_{g}(r)&={{g^{p/2}_{xx}(r)}}
\(1-\a(p-2)\frac{\[g_{{tt}}(r)g_{{xx}}^{{(p-3)}/{2}}(r)\]'g'_{xx}(r)}
{\[g_{tt}(r)g_{rr}(r)\]{g^{(p-1)/2}_{xx}(r)}}\).
\end{split}
\end{equation}
In (\ref{hFr}), one integration constant is the upper limit of the
integration, which is chosen to be  $r_c$ via the Dirichlet boundary
condition at the cutoff surface. The other integration constant is
chosen to be $\mG_{g}(r_h)$ to cancel out the first order zero in
the denominator ${\sqrt{g_{tt}(r_h)}}/{\sqrt{g_{rr}(r_h)}}=0$ of the
integrand in (\ref{hFr}), where
\begin{equation}\label{horizon}
\begin{split}
\mG_{g}(r_h)&\equiv\lim_{r\rightarrow r_h}\mG_{g}(r)={g^{p/2}_{xx}(r_h)}\(1-\a\Lambda_{\a}\),
~~~\Lambda_{\a}=\frac{4(p-2)\pi T_{H}}{\sqrt{g_{tt}(r_h)g_{rr}(r_h)}}\frac{g'_{xx}(r_h)}{{g_{xx}(r_h)}},
\end{split}
\end{equation}
and $\sqrt{g_{tt}(r_h)g_{rr}(r_h)}$ is a finite constant due to our
definition. The Hawking temperature $T_{H}$ is given in (\ref{TH}),
so that $\L_{\a}$ is a constant determined by the metric (\ref{gm1})
at the horizon.

With the perturbed metric $d\ts^2_{(0)}+d\ts^2_{(1)}$, corresponding
\BY~tensor becomes $\hT^{(0)}_{ab}+\hT^{(1)}_{ab}$, where the
symmetric traceless components of $\hT^{(1)}_{ab}$ are
\begin{equation}\label{BY1}
\begin{split}
\hT^{(1)}_{ij}&=-2\frac{\mG_g(r_h)}{g^{p/2}_{xx}(r_c)}\[1+\frac{\sqrt{g_{tt}(r_c)}}{\sqrt{g_{rr}(r_c)}}\hF'(r_c)\]
\si_{ij}
=-2\frac{\mG_g(r_h)}{g^{p/2}_{xx}(r_c)}\si_{ij}\equiv-2\te(r_c)\si_{ij}.
\end{split}
\end{equation}
Here $\te(r_c)$ is defined as the kinetic shear viscosity of fluid
dual to the  Gauss-Bonnet gravity with the geometry (\ref{gm1}).
Using the entropy density $s_{_0}(r_c)$ given in (\ref{s0}) of the
fluid at the cutoff surface, we have
\begin{equation}\label{shear}
\te(r_c)=\frac{\mG_g(r_h)}{g^{p/2}_{xx}(r_c)}
\Rightarrow~\frac{\te(r_c)}{s_{_0}(r_c)}=\frac{1}{4\pi}\[1-\a\Lambda_{\a}\].
\end{equation}
One can see that the shear viscosity over entropy density of dual
fluid does not run with the cutoff surface. This match with the
results in \cite{Iqbal:2008by} and \cite{Eling:2011ct}, where a
hypersurface in the given solution is introduced. Here we impose the
Dirichlet boundary conditions \cite{Bredberg:2010ky,Cai:2011xv} and
find the corresponding perturbed bulk solutions with a flat induced
hypersurface.  Using the formula (\ref{shear}), one can easily
obtain the ratio of shear viscosity over entropy density for some
 fluid once the dual gravity background is known. Take an example,
using the charged AdS Gauss-Bonnet black brane background solutions,
one has~\cite{Niu:2011gu,Cvetic:2001bk}
\begin{align}\label{AdS}
g_{xx}(r)=\frac{r^2}{\ell^2},
~~&g_{tt}(r)=g^{-1}_{rr}(r)=\frac{r^2}{2\l_G}\(1-\sqrt{1-\frac{4\l_G}{\ell^2}\[1-(1+Q^2)\frac{r_h^{p+1}}{r^{p+1}}+Q^2\frac{r_h^{2p}}{r^{2p}}\]}\),\nn\\
\Longrightarrow~~\frac{\te(r_c)}{\ts(r_c)}&=\frac{1}{4\pi}\[1-2\frac{\l_G}{\ell^2}\(\frac{p+1}{p-1}-Q^2\)\],
\qq \l_G=(p-1)(p-2)\a,
\end{align}
where $\alpha$ is the Gauss-Bonnet coefficient, and $\ell$ is the
radius of AdS  spacetime. When $\ell\rightarrow\infty$, the negative
cosmological constant $\L=-p(p+1)/2\ell^2\rightarrow 0$, the black
brane solution will degenerate into the ingoing Rindler space via
some coordinate redefinition~\cite{Bredberg:2010ky}, and the
Gauss-Bonnet term will not contribute to the shear viscosity any
more. This has been shown via using the ingoing Rindler metric of
flat spacetime~\cite{Chirco:2011ex}. Here our formula in
(\ref{horizon}) leads to this result directly because
$g_{xx}(r)=1\Rightarrow\L_\a=0$ .


\section{Holographic forced fluid at AdS boundary}
\label{sect:boundary}

It is interesting to give the concrete expressions of some external
forces of holographic fluid. For this, let us consider
 the Einstein-Maxwell action with a dilaton field
$\Phi$ and appropriate boundary terms and counter terms as follows.
\begin{equation}\label{eindil}
\mathcal {I}=\frac{1}{16 \pi G_{p+2}} \int d^{p+2}x \sqrt{-g}\left( R -2 \Lambda -\frac{1}{4}F^2 -\frac{1}{2}
(\na\Phi)^2\right)+S_{b.t.}+S_{c.t.},
\end{equation}
where the negative cosmological constant
$\Lambda=-{p(p+1)}/{2\ell^2}$, the Maxwell field $F_{MN}=2\na_{[M}
A_{N]}$, and $F^2=F_{MN}F^{MN}$,~$(\na\Phi)^2=\na_M \Phi \na^M
\Phi$~, where $\{M,N,...\}$ run over the bulk coordinates. In what
follows, $16 \pi G_{p+2}=1$ will be adopted. From the action we can
get the gravitational field equations
\begin{eqnarray}
\begin{aligned}
 R_{MN}-\frac{g_{MN}}{2}R+\Lambda g_{MN}&=\hf T^{(A)}_{MN}+\hf T^{(\Phi)}_{MN},
\end{aligned}
\end{eqnarray}
where
$T^{(A)}_{MN}=F_{MP}F^{~P}_{N}-\frac{g_{MN}}{4}F^2,
~T^{(\Phi)}_{MN}=\na_M \Phi \na_N \Phi -\frac{g_{MN}}{2}(\na \Phi)^2,
$
and from which one has
\begin{equation}
\begin{split}\label{gravity}
W_{MN}&\equiv R_{MN} + \frac{p+1}{\ell^2}
g_{MN}-\frac{1}{2}\partial_M \Phi \partial_N \Phi
-\frac{1}{2}\(F_{MP}F^{~P}_{N}-\frac{1}{2p}g_{MN}F^2\)=0.
\end{split}
\end{equation}
The equations of motion of the Maxwell field and the dilaton
field are
\begin{eqnarray}
\begin{aligned}
\label{gauge}
W_{N}&\equiv \nabla_{M}F^{M}_{~N}=0, \qq
W_{\phi}&\equiv \nabla^2 \Phi=0.
\end{aligned}
\end{eqnarray}
The system has a class of exact solutions with $p+2$ parameters as
\begin{eqnarray}
\left\{
\begin{aligned}\label{constant}
ds^2& = -2 u_\mu dx^\mu dr +\frac{r^2}{\ell^2}\[ V(r) u_\mu u_\nu dx^\mu dx^\nu
+ {\cal P}_{\mu\nu}\] dx^\mu dx^\nu ,\\
\mA&=-U(r)u_\mu dx^\mu ,\\
\Phi&=\Phi_0.~~~~
\end{aligned}
 \right.
\end{eqnarray}
where $\{\m,\n,...\}$ run over the boundary coordinates.  This is a
charged boosted black brane solution in Eddington-Finkelstein
coordinates with a constant dilaton $\Phi_0$, and a constant unit
normalized velocity $u_\mu$,
\begin{equation}\label{unu}
u_\m={\g_{v}(-1,~v_i)},\qq
\g_{v}=\(1-v^2\)^{-\frac{1}{2}},\qq
v^2=v_iv^i=\d_{ij}v^iv^j.
\end{equation}
and $\eta^{\mu\nu}u_\mu u_\nu =-1$. In the metric of the solutions
(\ref{constant}),
\begin{equation}
\V(r)=1-\frac{m\ell^2}{r^{p+1}}+\frac{q^2\ell^2}{r^{2p}}\equiv1-(1+Q^2)\frac{r_h^{p+1}}{r^{p+1}}+Q^2\frac{r_h^{2p}}{r^{2p}}.
\end{equation}
Here the notations in \cite{Banerjee:2008th} have been used,
\begin{equation}\label{V0}
\begin{split}
V(r_h)&=
1-M+Q^2=0,
\qquad M\equiv\frac{m\ell^2}{r_h^{p+1}},\qquad Q\equiv\frac{q\ell}{r_h^{p}},
\end{split}
\end{equation}
where $r=r_h$ is the horizon location of the charged black brane
solution. The Hawking temperature of the black brane is given by
\begin{equation}\label{Hawking}
\begin{split}
T_H&=\left. \frac{1}{4\pi}\[\frac{r^2V(r)}{\ell^2}\]'\right
|_{r=r_h}=\[(p+1)M-2p~Q^2\]\frac{r_h}{4\pi\ell^2},
\end{split}
\end{equation}
In the gauge field part, we have chosen the gauge that $A_r=0$ and
\begin{eqnarray}
\begin{aligned}
U(r)=\m_{0}\(b_{0}-\frac{r_h^{p-1}}{r^{p-1}}\),\qq~\m_{0}=
\frac{q}{r_h^{p-1}}\sqrt{\frac{2p}{p-1}},
\end{aligned}
\end{eqnarray}
where $b_0 \m_0$ is the value of the gauge potential at the AdS
boundary. Usually $b_{0}=1$ is required to get a well defined  gauge
field  at the horizon \cite{Chamblin:1999tk,Chamblin:1999hg}, but
this is not necessary in the perturbative fluid/gravity
correspondence \cite{Erdmenger:2008rm}.
\subsection{Charged fluid in non-relativistic limit}
\label{sect:boundary1}
To perturb the metric (\ref{constant}), we take the same procedure
via regarding the parameters in the metric and gauge field as
functions of boundary coordinates $u_{\m}\rightarrow u_{\m}(x),~
r_h\rightarrow r_h(1+P(x))~$\footnote{We briefly denote the function
variables $(x^\m)$ as $(x)$ and some of them in the following
equations will be ignored.}. Without loss of generality, we take the
associated replacements as $m\rightarrow m\(1+P(x)\)^{p+1}$ and
$q\rightarrow q\(1+P(x)\)^{p}$ to keep $M$ and $Q$ as two constants.
The scalar field is replaced as $\Phi\rightarrow\phi(x^\m)$. Then
the solutions (\ref{constant}) will not solve the equations of
motion (\ref{gauge}) and (\ref{gravity}) any more. Using the so-called BMW limit at the boundary
\cite{Bhattacharyya:2008kq,Bredberg:2011jq}
\begin{equation}\label{nr}
\begin{split}
\p_ r \sim \e^0,\qq ~\p_ {i}\sim v_i\sim \p_i\phi \sim \e^1,\qq
\p_\t\sim P\sim\e^2,
\end{split}
\end{equation}
where we have added an assumption that $ \p_i\phi \sim \e^1$, we can
solve the equations of motion order by order in the small parameter
$\e$~\cite{Compere:2011dx,Cai:2011xv}. Actually the three parameters
$(m,~q,~r_h)$ relate to each other via $V(r_h)=0$, so that two of
them are independent parameters, for example, one can take $q$ and
$r_h$ as two independent parameters. We focus on the forced NS
equations in this paper, so $\d m\sim \d q\sim \d r_h \sim \e^2$
have been assumed in the non-relativistic limit. In the boundary
derivative expansion of the relativistic fluid, the solutions of the
Einstein-dilaton system \cite{Bhattacharyya:2008ji}, and the
Einstein-Maxwell system \cite{Erdmenger:2008rm,Banerjee:2008th}
 have been obtained  up to second order of the derivative expansion parameter.
We can extract the relevant terms up to the second order in our non-relativistic expansion parameter $\e$.
To compare with their results, we write our the non-relativistic perturbative solutions up to $\e^2$ in a covariation form,
\begin{eqnarray}
\left\{
\begin{aligned}\label{solution1}
ds^2 =& -2 u_\mu(x) dx^\mu dr + \frac{r^2}{\ell^2} \[-\V\(\kr\) u_\mu(x) u_\nu(x) +  \P_{\mu\nu}(x) \]dx^\mu dx^\nu \\
& + \frac{r^2}{\ell^2}\[ \ell^2 \mK(r)\th_{\phi}u_\mu u_\nu+ 2 \ell \mF(r) \sigma_{\mu\nu}+ \ell^2 \mF_{\phi}(r)(\s_\phi)_{\mu\nu}\]dx^\mu dx^\nu,\\
\mA=& A_\m(r,x) dx^\m= \[A^{(q)}_\m(r,x) +A^{(ex)}_\m(x)\] dx^\mu ,\qq \\
\Phi = &\phi(x) + \ell \mF(r) \(u^\mu\partial_\mu \phi\)  + \ell^2\mF_{\phi}(r)(\p^\m\p_\m\phi).
\end{aligned} \right.
\end{eqnarray}
Although higher order terms such as $u_\t u_i\sim\e^3$ are reserved,
they would not contribute to the calculation up to order $\e^2$.
In (\ref{solution1}), 
$\P_{\m\n}(x)=\eta_{\m\n}+u_\mu(x) u_\nu(x),~ \kr=\[1-P(x)\]r,$
and $A^{(q)}_\m(r,x)=-\[1+P(x)\]U(\kr)u_\mu$ 
is the perturbed and boosted potential. $A^{(ex)}_\m$ is an extra
electromagnetic field, which would provide extra forces and
$F^{(ex)}_{\m\n}=2\p_{[\m} A^{(ex)}_{\n]}$ are chosen to meet with
the appropriate non-relativistic expansion in
\cite{Bredberg:2010ky}, $F^{(ex)}_{\t
i}\sim\e^3,~F^{(ex)}_{ij}\sim\e^2$, as well as the velocity $
u_\m=(-1- v^2/2 , ~v_i)$.
 This solution  solves the equations of
motion (\ref{gauge}) and (\ref{gravity}) up to order $\e^2$,
 if the following expressions are provided
\begin{eqnarray}
\begin{aligned}\label{shear1}
\mF(r)=\ell\int_{r}^\infty  \frac{\(y^{p}-r_h^{p}\)}{y^{p+2} V(y)}dy,\qq
\sigma_{\mu\nu} =\p_{(\m} u_{\n)}-\frac{\P_{\mu\nu}}{p}\th,
\end{aligned}
\end{eqnarray}
where $\th=\eta^{\m\n}\p_\m u_\n=\P^{\m\n}\p_\m u_\n$, for the
dilaton shear perturbations
\begin{eqnarray}
\begin{aligned}\label{shear2}
\mF_{\phi}(r)=\frac{\ell^2}{p-1}\int_{r}^\infty \frac{(y^{p-1}-r_h^{p-1})}{y^{p+2}V(y)}dy,\qq
(\s_\phi)_{\m\n}=\p_{(\m} \phi\p_{\n)}\phi - \frac{\P_{\mu\nu}}{p}\th_{\phi},
\end{aligned}
\end{eqnarray}
and for the dilaton scalar perturbations
\begin{eqnarray}
\begin{aligned}
\qq\mK(r)=\frac{\ell^2}{2p(p-1)r^2},\qq\qq
\th_{\phi}=\P^{\m\n}\p_\m\phi\p_\n\phi.
\end{aligned}
\end{eqnarray}
We can generalize the dual boundary fluid
in~\cite{Bhattacharyya:2008ji,Banerjee:2008th} to the (p+1)-dimensional case with $p\geq2$, where the stress tensor
$\hT^{\mu}_{\nu}$ and current $\hJ^{\m}$ are given by
\begin{eqnarray}
\begin{aligned}\label{boundary}
\mT^\mu_\nu &= \lim_{r\rightarrow\infty} \frac{r^{p+1}}{\ell^{p+1}}\left[2(K_{\alpha\beta}h^{\alpha\beta} h^\mu_\nu - K^\mu_\nu)
-\frac{2p}{\ell} h^\mu_\nu +\frac{2\ell}{p-1}\hG^\mu_\nu -\frac{\ell}{p-1}(\hG_{\Phi})^\mu_\nu \right],\\
\mJ^\m&=\lim_{r\rightarrow\infty} \frac{r^{p+1}}{\ell^{p+1}} N_{r}F^{\m r},
~~~~\hG^\mu_\nu=\hat{R}^\mu_\nu-\frac{1}{2}\hat{R} h^\mu_\nu,
~~~(\hG_{\Phi})^\mu_\nu=\hn^\mu \Phi\hn_\nu\Phi-\frac{1}{2} (\hn \Phi)^2 h^\mu_\nu, \\
\end{aligned}
\end{eqnarray}
where $N$ is the outward pointing unit normal of the regulated boundary,
$\hn$ is the covariant derivative associated with the boundary
metric $h^\m_{\n}$, and the corresponding Einstein tensor
$\hG^\mu_\nu$, as well as the effective counter tensor
$(\hG_{\Phi})^\mu_\nu$ of the dilaton field \cite{Kraus:1999di,Bianchi:2001kw,Skenderis:2002wp}.
Substituting our solutions (\ref{solution1}) in (\ref{boundary}),
the Brown-York stress tensor and the induced current at the AdS
boundary are given by
\begin{eqnarray}
\begin{aligned}\label{Tmn}
 \mT^{\mu\nu}&=M \frac{ r_h^{p+1}}{\ell^{p+2}}[1+(p+1)P]
\left[(p+1)\, u^\mu u^\nu + \eta^{\mu \nu} \right] -2\,  \frac{r_h^{p}}{\ell^{p}} \,\sigma^{\mu \nu}  -
\frac{1}{p-1} \frac {r_h^{p-1}}{\ell^{p-2}} \,(\sigma_{\phi})^{\mu\nu},\\
\mJ^{\m}&=\frac{r^{p}}{\ell^{p}}F^{\m r}=n_{(q)} u^\m,~~~~
n_{(q)}=(1+pP)n_0,~~~~ n_0 \equiv \frac{r^p}{\ell^p}U'(r) =\frac{q\sqrt{ 2p(p-1)}}{\ell^p},
\end{aligned}
\end{eqnarray}
where $n_{(q)}$ is the induced charge density. Note that here
$(\s_{\phi})^{\m\n}$ is included in the stress tensor (\ref{Tmn}).
These terms relating to the dilaton field were considered as the
dissipative parts of the dual fluids' stress tensor in
\cite{Bhattacharyya:2008ji}, or the perturbations in the membrane
paradigm \cite{Iqbal:2008by}. In this paper, however, we focus on
the forced NS equations and move them to the right-hand
side of the constraint equations as additional sources, similar to
what has been done for the boundary metric perturbations
~\cite{Bhattacharyya:2008kq,Brattan:2011my}. Then we can redefine
the stress tensor of the dual fluids with first order dissipative
term
\begin{align}\label{TNS}
 \mT_{\mu\nu}^{(\o)}=&~\o_0\[1+(p+1)P\]
\left[\, u_\mu u_\nu+\frac{\eta_{\mu \nu}}{p+1}\right] -2\, \eta_0  \,\sigma_{\mu \nu},
\qq\o_0=(p+1)M\frac{ r_h^{p+1}}{\ell^{p+2}},
\end{align}
where $\eta_0$ is the holographic shear viscosity of the dual fluid.
With the corresponding entropy density $s_0$, we have
\begin{equation}
\begin{split}
\eta_0=\frac{r_h^{p}}{\ell^{p}},\qq
s_0=\frac{1}{4G_{p+2}}\frac{r_h^p}{\ell^p}~\Rightarrow~
\frac{\eta_0}{s_0}=\frac{1}{4\pi},
\end{split}
\end{equation}
and the thermodynamic relation $T_Hs_0=\o_0-n_0\m_0$ still holds.
Furthermore, we can redefine the following stress tensor by moving
out the terms associated with the boundary chemical potential
\begin{equation}\label{Tabs}
\begin{split}
 \mT_{\mu\nu}^{(s)}&={T_Hs_0}\[1+(p+1)P\]
\left[\, u_\mu u_\nu+\frac{\eta_{\mu \nu}}{p+1}\right]  -2\, \eta_0  \,\sigma_{\mu \nu},\\
\end{split}
\end{equation}
This form will be used later.
\subsection{Forced Navier-Stokes equations}
\label{sect:boundary2}
Using Gauss-Codazzi-Mainardi relations near the
boundary~\cite{Bhattacharyya:2008ji}, we can obtain the constraint
equations of gravity
\begin{equation}\label{Gauss}
\begin{split}
&2D_A\(K_{CD} h^{CD} h^A_{B}-K^A_{B}\)=-2R_{CD}h^C_B N^D=-D_B \Phi \na_D \Phi~ N^D- F_{CP}F^{P}_{~~D}h^C_B~N^D,
\end{split}
\end{equation}
where 
$h_{AB}$,~$N^A$ and $K_{AB}$ are respectively the  induced metric,
the outward pointing unit normal and the extrinsic curvature of the
constant $r$ hypersurface, with the covariant derivative $D$. When
$r\rightarrow\infty$, transforming these results into the boundary
metric $\eta_{\m\n}$ of the dual fluid, we have
\begin{equation}\label{conser}
\p_\mu \mT^{\mu}_{\nu}=-\lim_{r\rightarrow\infty} \frac{r^{p+1}}{\ell^{p+1}}\[\left(N^C\na_C\Phi+\frac{\ell}{p-1}\hn^2\Phi\right)\hn_\n\Phi
+ N^CF_{C D}F^{~D}_{\n}\].
\end{equation}
Introduce the gauge field tensor $F^{(bd)}_{\m\n}=2\p_{[\m}
A^{(bd)}_{\n]}$ which comes from the boundary chemical potential
$A^{(bd)}_\m=-b_0\m_{0}(1+P)u_\m~$, and denote the background gauge
field tenor as $F^{(bg)}_{\m\n}=2\p_{[\m} A^{(bg]}_{\n)}$, where
$A^{(bg)}_\m=A^{(bd)}_\m+A^{(ex)}_\m~$, the $(p+1)$ constraint
equations up to $\e^3$ become
\begin{align}\label{costrain1}
&\p^\mu \mT_{\mu \nu}=-\p_\nu \phi\[\frac{r_h^p}{\ell^{p}} ~(u^\mu\partial_\mu \phi)
+ \frac{1}{p-1} \frac{r_h^{p-1}}{\ell^{p-2}}(\p^\m\p_\m\phi )\]- \mJ^{\m}F^{(bg)}_{\m\n},
\end{align}
where $F^{(bg)}_{\m\n}=F^{(ex)}_{\m\n}+F^{(bd)}_{\m\n}$, and the
constraint equation of Maxwell field reads $\p^\mu
\mJ_\m=n_{(q)}\p^\m u_\m=0$. Using the redefined stress energy
tensor (\ref{TNS}), we get the forced NS equations at order $\e^3$,
\begin{align}\label{NSomega}
~~~\p^\mu \mT_{\mu \nu}^{(\o)}&=f_\n^{(\phi)}+f_\n^{(q)},\qq
 f^{(q)}_{\n}=-n_{(q)}u^\m F^{(bg)}_{\m\n},
\end{align}
where $f^{(q)}_{\n}$ is just the Lorentz force of the charged fluid
and $f^{(\phi)}_{\nu}$ is the external force from the dilaton field
up to $\e^3$
\begin{eqnarray}
\begin{aligned}
f^{(\phi)}_{\nu}\equiv&-\chi_0\( u^\mu\partial_\mu \phi \p^\nu\phi\)
 +\xi_0\(\p^\m\phi\p_\m\p^\n\phi\),
~~~\chi_0=\frac{r_h^p}{\ell^{p}}, ~~~\xi_0=\frac{p-2}{p(p-1)}
\frac{r_h^{p-1}}{\ell^{p-2}}.
\end{aligned}
\end{eqnarray}
The Lorentz force up to $\e^3$ can also be divided into two parts as
\begin{eqnarray}
\begin{aligned}
f^{(q)}_{\nu}\equiv& 
 f^{(ex)}_{\n}+f^{(bd)}_{\n},
\qq~~ f^{(ex)}_{\n}=-n_{(q)}u^\m F^{(ex)}_{\m\n},
\qq~~ f^{(bd)}_{\n}=-n_{(q)}u^\m F^{(bd)}_{\m\n}.
\end{aligned}
\end{eqnarray}
In addition, we notice that  the following relation holds at order
$\e^3$
\begin{align}\label{TNS2}
\p^\m [\mT_{\mu\nu}^{(\o)}-\mT_{\mu\nu}^{(s)}]b_0=f^{(bd)}_{ \n}
~~\xrightarrow[]{b_0=1}~~\p^\mu \mT_{\mu
\nu}^{(s)}=f_\n^{(\phi)}+f_\n^{(ex)}.
\end{align}
Thus, if we define the fluid stress energy tensor as (\ref{Tabs}),
with $b_0=1$, we can see that the external forces only come from the
dilaton field and the external electromagnetic field. This result is
consistent with the one in \cite{Bhattacharyya:2008kq}. It would be
more clear to see the non-relativistic results by neglecting higher
order terms of the solutions in the covariant form. The solutions in
(\ref{solution1}) become
\begin{eqnarray}
\left\{
\begin{aligned}\label{dsaphi}
ds^2 =&+ 2 d\t dr + \frac{r^2}{\ell^2} \[-\V\(r\)d\t^2 +  \d_{ij}dx^i dx^j \]   -2v_i   dx^i dr -2\frac{r^2}{\ell^2}\[1-\V(r)\]v_idx^i d\t\\
& + v^2 d\t dr+  \frac{r^2}{\ell^2} \[r\V'(r)P d\t^2+  \(1-\V\(r\)\)\(v^2 d \t^2+ v_i v_j dx^i dx^j\)\] \\
& + \frac{r^2}{\ell^2}\[\ell^2\mK(r) \theta_{\phi} d\t^2+ \(2\ell\mF(r) \sigma_{ij}+ \ell^2\mF_{\phi}(r)(\s_\phi)_{ij}\)dx^i dx^j\],\\
\mathcal {A}=&\[U(r)\(1+P+v^2/2\)-rP U'(r)+A^{(ex)}_\t\] d\t+\[-U(r)v_i+A^{(ex)}_i\] dx^i,\\
\Phi = &\phi(x)  +\ell \mF(r) \(\partial_\t \phi+v^i\partial_i \phi\)  + \ell^2 \mF_{\phi}(r)(\p^i\p_i\phi), \\
\end{aligned} \right.
\end{eqnarray}
up to $\e^2$, where the shear tensors are
\begin{equation}\label{scaltwoderi}
\begin{split}
\sigma_{ij} &=\p_{(i} v_{j)}- \frac{\d_{ij}}{p}\th,
\qq(\sigma_\phi)_{ij}=\p_{(i} \phi\p_{j)}\phi- \frac{\d_{ij}}{p}\th_\phi,
\end{split}
\end{equation}
and $\th=\d^{ij}\p_i v_j$,~$\th_\phi=\d^{ij}\p_i\phi\p_j\phi$~. Our
aim is to obtain the forced NS equations, which turn out to be the
constraint equations of the gravitational field equations
(\ref{gravity}) at order $\e^3$. The solutions up to $\e^2$ are
enough to provide the forced NS equations, because higher order
corrections do not make contribution  in this order.
 With the solutions (\ref{dsaphi}), the constraint equations at order $\e^2$ give the incompressible
condition $\o_0\p_iv^i=0$, and at order $\e^3$ give us with
\begin{equation}\label{FNS} \begin{split}
\o_0\(\p_\t v_i + v^j\p_j v_i+\p_i P \)-\eta_0\p^2 v_i&=f^{(\phi)}_i+f^{(q)}_i.\\
\end{split}
\end{equation}
These equations are just the temporal and spatial components of the
equations in (\ref{NSomega}) up to $\e^3$. Here  the external forces
only have the spatial components as
\begin{align}
f^{(\phi)}_i= -\chi_0\(v^j\p_j\phi\p_i\phi\)+\xi_0\(\p^j\phi\p_j\p_i\phi\),
\qq f^{(q)}_{i}= - n_0\(E^{(bg)}_{i}+v^jB^{(bg)}_{ji}\),
\label{forceq}
\end{align}
where $ E^{(bg)}_{i}=F^{(bg)}_{\t i}$ and
$B^{(bg)}_{ij}=F^{(bg)}_{ij}$ are the background electric and
magnetic fields respectively. Further we define the dynamic
viscosity and the normalized forces as
\begin{equation}
\n_{[\o]}={\eta_0}/{\o_0},\qq
{f}^{(\phi)}_{[\o]i}={f}^{(\phi)}_i/{\o_0},\qq~{f}^{(q)}_{[\o]i}={{f}^{(q)}_i}/{\o_0},
\end{equation}
the forced incompressible NS equations then become
\begin{equation}\label{f1}
\begin{split}
&\p_\t v_i + v^j\p_j v_i+\p_i P  -\n_{[\o]}\p^2 v_i= {f}^{(\phi)}_{[\o]i}+{f}^{(q)}_{[\o]i},\qq \p_i v^i=0.\\
\end{split}
\end{equation}
 If we consider the characteristic scale $L\sim\e^{-1}$ and the velocity $v=\sqrt{v_iv^i}\sim\e$,
the Reynolds number of the dual fluid
 $\mR_e={vL}/{\nu_{[\o]}}\sim 4\pi T_H\[1+(n_0\m_0)/(T_Hs_0)\]$.

In addition, we can  divide the electromagnetic forces into two
parts $f^{(q)}_i=f^{(bd)}_i+f^{(ex)}_i$, where
\begin{align}
f^{(ex)}_{i}=&- n_0\(E^{(ex)}_{i}+v^jB^{(ex)}_{ji}\),\qq
f^{(bd)}_i = ~ b_0 n_0  \m_0 (\p_\t v_i+ v^j \p_j v_i+\p_i P),
\end{align}
where $ E^{(ex)}_{i}=F^{(ex)}_{\t i}$ and
$B^{(ex)}_{ij}=F^{(ex)}_{ij}$. Then $f_i^{(ex)}$ is just the Lorentz
force from the extra electromagnetic field, while $f^{(bd)}_i$ from
the boundary chemical potential. We now move the term $f_i^{(bd)}$
to the left-hand side of the NS equations. Defining the dynamic
shear viscosity and the force density as
\begin{equation}
\begin{split}
\nu_{[s]}=\frac{\eta_0}{T_H s_0}=\frac{1}{4\pi T_H},\qq
{f}_{{[s]}i}^{(\phi)}= \frac{{f_i^{(\phi)}}}{T_H s_0},\qq
{f}_{{[s]}i}^{(ex)}=\frac{{f_i^{(ex)}}}{T_H s_0},
\end{split}
\end{equation}
we can rewrite the incompressible charged NS equations  as
\begin{equation}\label{NS2}
\begin{split}
\p_{\tau} v_i + v^j\p_j v_i+\p_i P-\nu_{[s]}\p^2 v_i
={f}_{{[s]i}}^{(\phi)}+ {f}_{{[s]}i}^{(ex)},\qq \p_iv^i=0.
\end{split}
\end{equation}
In this case, the Reynolds number becomes
$\mR_e={vL}/{\nu_{[s]}}\propto T_H$, proportional to the temperature
of the fluid.

\section{Holographic forced fluid at cutoff surface}
\label{sect:cutoff}

In this section, we will generalize the previous discussions to the
case of dual fluid at a finite cutoff surface by using the method
which is introduced in Section \ref{sect:review}. In this case, we
only need to substitute
\begin{equation}
g_{tt}(r)=g^{-1}_{rr}(r)={r^2}V(r)/{\ell^2},\qq
g_{xx}(r)={r^2}/{\ell^2},
\end{equation}
into the generic metric (\ref{gm1}), and introduce a finite cutoff
surface in the background solution and consider the non-relativistic
expansions
\begin{equation}\label{nr2}
\begin{split}
\pc_ r \sim \e^0,\qq ~\pc_ {i}\sim \b^i\sim \pc_i\phi \sim \e^1,\qq
\pc_\t\sim \tP\sim\e^2.
\end{split}
\end{equation}
The equations of motion to be solved are given by (\ref{gauge}) and
(\ref{gravity}).

\subsection{Charged fluid in non-relativistic limit}
\label{sect:cutoff1}

Since we are considering the bulk solution with a finite cutoff in
this section, following \cite{Brattan:2011my}, we  choose the gauge
that $g_{rr}=0$, $g_{ra}\propto\u_a$ and $g^{_{(1)}}_{ii}=0$. The
perturbed solution up to order $\e^2$ with a Dirichlet boundary
condition at the cutoff surface turns out to be
\begin{eqnarray}
\left\{
\begin{aligned}\label{solution}
d\ts^2 =& -\frac{2\ell\[1+\hH(\kr)(r_c^2\tth_{\phi})\]}{\kr_c~\sqrt{V(\kr_c)}} \u_a(\x) d\x^ad\kr + \frac{\kr^2}{\kr_c^2} \[-\frac{V\(\kr\)}{V\(\kr_c\)} \u_a(\x) \u_b(\x) +  \Pc_{ab}(\x) \]d\x^a d\x^b \\
&+ \frac{r^2}{r_c^2} \[ \hK(r)(r_c^2\tth_{\phi})\frac{\u_a \u_b}{V(r_c)}+2{r_c~} \hF(r){ \sqrt{V(r_c)}}~\si_{ab}+ r_c^2\hF_{\phi}(r)(\si_\phi)_{ab}\]d\x^a d\x^b,\\
\hA=&\A_a(\kr,\x)d\x^a= \[\A^{(bd)}_a(\kr,\x)+\A^{(ex)}_a(\x)\] d\x^a
+\frac{\ell}{r_c}\frac{\m_0\hQ(r) }{\sqrt{V(r_c)}}\frac{r_h^{p-1}}{r^{p-1}} (r_c^2\tth_{\phi}) \u_a(\x) d\x^a,\\
\tPh = &\phi(\x^a) + r_c \hF(r)\sqrt{V(r_c)} \(\u^a\pc_a \phi\)  + r_c^2\hF_{\phi}(r)\(\pc^a\pc_a\phi\), \\
\end{aligned} \right.
\end{eqnarray}
where the gauge field relates to (\ref{solution1}) through
\begin{align}
\A^{(bd)}_a(\kr,\x)&=-\frac{\ell}{\kr_c}
\frac{U(\kr)}{\sqrt{V(\kr_c)}} \u_a(\x),~~~
\A^{(ex)}_{a}(\x) =-\frac{\ell }{\kr_c}\frac{A^{(ex)}_\t(x)}{\sqrt{V(\kr_c)}}\u_a(\x)
+\frac{\ell }{\kr_c~}A^{(ex)}_i(x)\tnq^i_a(\x).
\end{align}
and $\tnq^i_a(\x)$ are given in (\ref{bt}) with a coordinate-dependent velocity.
Again keep in mind that we have already taken
the non-relativistic limit (\ref{nr2}) of the solution
(\ref{solution}) up to order $\e^2$ and higher order terms do not
make contribution in the following calculations, they are reserved
just for a covariant form. For the shear perturbations due to the
boost,
\begin{eqnarray}
\begin{aligned}\label{scal2}
\hF(r)&=\ell \int_{r}^{r_c}  \frac{\(y^{p}-r_h^{p}\)}{y^{p+2} V(y)}dy,
\qq~~~\si_{ab} =\pc_{(a} \u_{b)}-\frac{\Pc_{ab}}{p}\tth.
\end{aligned}
\end{eqnarray}
and $\tth=\te^{ab}\pc_a \u_b\equiv\Pc^{ab}\pc_a \u_b$. For the shear
perturbations due to the dilaton field,
\begin{eqnarray}
\begin{aligned}
\hF_{\phi}(r)&=\frac{\ell^2}{p-1}\int_{r}^{r_c}
\frac{(y^{p-1}-r_h^{p-1})}{y^{p+2}V(y)}dy,
\qq(\si_\phi)_{ab}=\pc_{(a} \phi\pc_{b)}\phi -
\frac{\Pc_{ab}}{p}\tth_{\phi},
\end{aligned}
\end{eqnarray}
where
$~\tth_{\phi}\equiv\Pc^{ab}(\pc_a\phi\pc_b\phi)$. 
The corresponding scalar perturbation equations give the following
solutions
\begin{eqnarray}
\begin{aligned}\label{scalar}
\hH(r)&=\frac{\ell^2}{4p(p-1)r_c^2}\frac{h(r_c)}{\sqrt{V(r_c)}},\qq
\hQ(r)=\frac{\ell^2}{4p(p-1)r_c^2}\frac{a(r_c)}{\sqrt{V(r_c)}}\(1-\frac{r^{p-1}}{r_c^{p-1}}\),\\
\hK(r)&=\frac{\ell^2}{2p(p-1)r^2}\[1-\frac{r^2}{r_c^2}\frac{h(r_c)V(r)}{\sqrt{V(r_c)}}+\frac{r_c^{p-1}}{r^{p-1}}\(h(r_c)\sqrt{V(r_c)}-1\)
+\tilde{q}(r)\].
\end{aligned}
\end{eqnarray}
Here two of the integration constants in (\ref{scalar}) have been
determined via the Dirichlet boundary condition of $\hK(r)$ and
$\hQ(r)$. The other two integration constants $a(r_c)$, $h(r_c)$,
and the notation
\begin{eqnarray}
\begin{aligned}
\tilde{q}(r)&=\frac{h(r_c)-a(r_c)}{\sqrt{V(r_c)}}\frac{r^2}{r_c^2}\frac{q^2\ell^2}{r^{2p}}
\(1-\frac{r^{p+1}}{r_c^{p+1}}\),
\end{aligned}
\end{eqnarray}
will be determined via choosing the Landau gauge of the stress
energy tensor of dual fluid and the induced charge current, whose
expressions are given by
\begin{eqnarray}
\begin{aligned}
\hT_{ab}(r_c)&=2(\K\eta_{ab}-\K_{ab})+\hT^{(ct)}_{ab},
\qq\hJ^{(q)}_a(r_c)=\N_{C}\F_a^{~C}.
\end{aligned}
\end{eqnarray}
The counter term at the cutoff surface is
\begin{align}
\hT^{(ct)}_{ab}&=- \cc(r_c)\[\frac{2p}{\ell}\te_{ab}+\frac{
\ell}{(p-1)}\(\pc_a\phi\pc_b\phi -\frac{\te_{ab}}{2}(\pc\phi)^2\)\],
\end{align}
where $\tilde {c}(r_c)$ could be an arbitrary function of $r_c$.  In
fact, at the finite cutoff surface, the counter term is not
necessary. To match the results at the AdS boundary in the case
without the cutoff surface, we here add the counter term and can
take $\tilde {c}(r_c)=1$.

The stress energy tensor and current in the derivative expansion of
the fluid dual to the perturbed solution (\ref{solution}) can be
written as
\begin{eqnarray}
\begin{aligned}\label{STJ}
\hT_{ab}(r_c)&=\hT_{ab}^{(0)}+ \hT_{ab}^{(1)}+\hT_{ab}^{(2)}+...,\qq
\hT_{ab}^{(0)}= \to(r_c)\u_a \u_b+\tp(r_c)\te_{ab},\\
\hJ^{(q)}_a(r_c)&=\hJ^{(0)}_a+\hJ^{(1)}_a+\hJ^{(2)}_a+...,~~~~~
\hJ^{(0)}_a=\tnq(r_c)\u_a,
\end{aligned}
\end{eqnarray}
where $\to(r_c)=\tr(r_c)+\tp(r_c)$. The zeroth order parts in
derivative expansion are just the stress energy tensor and current
of ideal charged fluid. The energy density $\tr(r_c)$, pressure
$\tp(r_c)$, and charge density $\tnq(r_c)$ up to order $\e^2$ in the
non-relativistic expansion are given by
\begin{align}\label{energy2}
\tr(r_c)&=\tr_{_0}(r_c)-(r_cP)\tr'_{_0}(r_c),
\qq\tr_{_0}(r_c)=-\frac{2p\sqrt{V(r_c)}}{\ell}+\frac{2p~\cc(r_c)}{\ell},\nn\\
\tp(r_c)&=\tp_{_0}(r_c)-(r_cP)\tp'_{_0}(r_c),
\qq\tp_{_0}(r_c)=\frac{r_cV'(r_c)}{\ell\sqrt{V(r_c)}}-\tr_{0}(r_c),\\
\tnq(r_c)&=\tnq_{_0}(r_c)-(r_cP)\tnq'_{_0}(r_c),
\qq\tnq_{_0}(r_c)=U'(r_c).\nn
\end{align}

The higher order terms in derivative expansion (\ref{STJ})  are the
dissipative parts of the stress energy tensor and current, we denote
them by $\hT^{(diss)}_{ab}$ and $\hJ^{(diss)}_{ab}$. Then using the
Landau gauge up to $\e^2$,
\begin{align}\label{landau}
u^a\hT^{(diss)}_{ab}=0\Rightarrow h(r_c)=\cc(r_c),
\qq\u^a\hJ^{(diss)}_a=0\Rightarrow a(r_c)=h(r_c),
\end{align}
so that $\tilde{q}(r)=0$, and when $r_c\ri \infty $ we can recover
the results (\ref{solution1}) at the AdS boundary  via transforming
them to the coordinates in the boundary.

 After imposing the Landau gauge
(\ref{landau}) up to order $\e^2$, we have
$\hJ_{a}^{(1)}=\hJ_{a}^{(2)}=0$  and
\begin{align}
\hT_{ab}^{(1)}&= - 2\te_{_0}(r_c)\si_{ab},\qq \hT_{ab}^{(2)}= -
2\te_{\phi}(r_c)(\si_{\phi})_{ab}-\tz_{\phi}(r_c)\tth_{\phi}\P_{ab},
\end{align}
where the cutoff-dependent shear viscosity and entropy density are
\begin{align}\label{tsrc}
\te_{_0}(r_c)=\frac{r_h^{p}}{r_c^{p}},\qq
\ts_{_0}(r_c)=\frac{1}{4G_{p+2}}\frac{r_h^{p}}{r_c^{p}}
~~\Rightarrow~~\frac{\te_{_0}(r_c)}{\ts_{_0}(r_c)}=\frac{1}{4\pi}.
\end{align}
And the coefficients associated with dilaton $\phi$ are
\begin{eqnarray}
\begin{aligned}
\te_{\phi}(r_c)&=\frac{\ell}{2(p-1)\sqrt{V(r_c)}} \[ \frac{r_h^{p-1}}{r_c^{p-1}} + \cc(r_c)\sqrt{V(r_c)}-1  \],\\
\tz_{\phi}(r_c)&=\frac{\ell}{2p \sqrt{V(r_c)}}\[1-\cc(r_c)\sqrt{V(r_c)}-\frac{ \cc(r_c) r_cV'{(r_c)}}{2(p-1)\sqrt{V(r_c)}}\].
\end{aligned}
\end{eqnarray}
On the other hand, at the zeroth order, from (\ref{TH}) and
(\ref{energy2}), one has
\begin{equation}
\begin{split}
\T_{_0}(r_c)=\frac{r_h^2 V'(r_h)}{4\pi \ell r_c\sqrt{V(r_c)}},
\qq{\to_{_0}(r_c)}=\tr_{_0}(r_c)+\tp_{_0}(r_c)=\frac{r_cV'(r_c)}{\ell\sqrt{V(r_c)}},
\end{split}
\end{equation}
as well as $\tnq_{_0}(r_c)$ in (\ref{energy2}), and $\ts_{_0}(r_c)$
in (\ref{tsrc}), thus one can show that the following
thermodynamical relation still holds for the  fluid at the cutoff
surface
\begin{equation}
\begin{split}
&\T_{_0}(r_c) s_{_0}(r_c)={\to_{_0}(r_c)}- \tnq_{_0}(r_c)\tm_{_0}(r_c),\qq\tm_{_0}(r_c)=\frac{\ell \[U(r_c)-U(r_h)\]}{r_c~\sqrt{V(r_c)}},
\end{split}
\end{equation}
where the chemical potential $\tm_{_0}(r_c)$ is defined as the difference
of the  gauge potential  between the horizon and the cutoff surface.
The dimensionless coordinate invariant diffusivity defined in \cite{Bredberg:2010ky}
is
\begin{equation}
\begin{split}
\bar{D}_c(r_c)=\T_{_{0}}(r_c)\frac{\te_{_{0}}(r_c)}{\to_{_{0}}(r_c)}
=\frac{1}{4\pi}\[1+\frac{\tnq_{_{0}}(r_c)\m_{_{0}}(r_c)}{\T_{_{0}}(r_c)
s_{_{0}}(r_c)}\]^{-1},
\end{split}
\end{equation}
which shows the dependence of the cutoff surface when the chemical
potential is present.

\subsection{Forced Navier-Stokes equations}
\label{sect:cutoff2}

As what has been done in section \ref{sect:boundary1}, we can
redefine the stress energy tensor of the dual fluid as
\begin{align}\label{Tabu}
\hT^{(\o)}_{ab}&=\to(r_c)\u_a\u_b+\tp(r_c)\tilde{\eta}_{ab}-2\te_{_0}(r_c)\si_{ab},
\end{align}
by moving out the part of dilaton field. By using the constraint
equations  at the cutoff surface for gravitational field
(\ref{Gauss}) and Maxwell field, up to $\e^3$ we can get the forced
NS equations and
 the conservation equation of charge current
\begin{align}\label{tNS1}
\p^a \hT_{ab}^{(\o)}& =\f^{(\phi)}_{b}(r_c)+\f^{(q)}_{b}(r_c),
\qq\p^a\hJ^{(q)}_a=\tnq_{_0}(r_c)\p^a\u_a=0,
\end{align}
where the external force from the dilaton field is
\begin{eqnarray}
\begin{aligned}
&\f^{(\phi)}_{b}(r_c)
=-\tch_{_0}(r_c)\(\u^a\pc_a\phi\pc_b\phi\)+\txi_{_0}(r_c)\(\pc^a\pc_a\phi\pc_b\phi\),
\qq\tch_{_0}(r_c)=\te_{_0}(r_c)=\frac{r_h^p}{r_c^p},\\
&\txi_{_0}(r_c)=\frac{(p-2)\ell}{p(p-1)\sqrt{V(r_c)}}
\[\frac{r_h^{p-1}}{r_c^{p-1}}+\frac{1}{(p-2)}\(1
-{\cc(r_c)}{\sqrt{V(r_c)}}-\frac{\cc(r_c)r_cV'{(r_c)}}{2\sqrt{V(r_c)}}\)\].
\end{aligned}
\end{eqnarray}%
And the Lorentz force of the charged fluid is
\begin{align}
\f^{(q)}_{b}(r_c)=\f^{(ex)}_{b}+\f^{(bd)}_{b},\qq
\f^{(ex)}_{b}=-\hJ^{a}_{(q)}\F^{(ex)}_{ab},\qq
\f^{(bd)}_{b}=-\hJ^{a}_{(q)}\F^{(bd)}_{ab},
\end{align}
with the  external Maxwell tensor
$\F^{(ex)}_{ab}=2\p_{[a}\A_{b]}^{(ex)}(r_c)$,~
and the induced Maxwell tensor
$\F^{(bd)}_{ab}=2\p_{[a}\A_{b]}^{(bd)}(r_c)$, where the gauge
potential on the cutoff surface is
\begin{align}
\A_{a}^{(bd)}(r_c)=-\[\tilde{\m}_{_{(b)}}(r_c)-(r_cP)\tilde{\m}'_{_{(b)}}(r_c)\]\u_{a},\qq
\tilde{\m}_{_{(b)}}(r_c)=\frac{\ell U(r_c)}{r_c\sqrt{V(r_c)}}.
\end{align}
 If we  furthermore redefine the following stress energy tensor
by moving out the terms associated with the boundary chemical
potential
\begin{align}\label{TNS3}
\hT_{ab}^{(s)}=\T_{_0}(\kr_c) s_{_0}(\kr_c)\u_a \u_b+
\[\tp_{_0}(\kr_c)-\tnq_{_0}(r_c)\tm_{_{0}}(\kr_c)\]\te_{ab}-2\te_{_0}(r_c)\si_{ab}.
\end{align}
where $\kr_c=r_c(1-P)$, when $b_0=1$, one has $U(r_h)=0$, and
$\tm_{_{(b)}}(r_c)$ is just the chemical potential $\tm_{_0}(r_c)$
of the fluid at the cutoff surface, the following relation holds up
to order $\e^3$,
\begin{align}\label{tTNS2}
\pc^a
[\hT_{ab}^{(\o)}-\hT_{ab}^{(s)}]\tm_{_{(b)}}(r_c)=\tm_{_{0}}(r_c)\f^{(bd)}_{
b} ~~\xrightarrow[]{b_0=1}~~\pc^a
\hT_{ab}^{(s)}=\f_b^{(\phi)}+\f_b^{(ex)}.
\end{align}
This keeps the same form as in the case at AdS boundary discussed in
the previous section. In this case, the external forces only come
from the dilaton field and the external electromagnetic field. To
see the above results more clearly, we write the them up to the
desired order. For example, the solutions (\ref{solution}) up to
order $\e^2$ are,
\begin{eqnarray}
\left\{
\begin{aligned}\label{dsphi}
d\ts^2 =&+ \frac{2\ell}{r_c~\sqrt{V(r_c)}}  d\t dr + \frac{r^2}{r_c^2} \[-\frac{V\(r\)}{V\(r_c\)}d\t^2 +  \d_{ij}dx^i dx^j \]-  \frac{r^2}{r_c^2}  \[1-\frac{V\(r\)}{V\(r_c\)}\]2\b_idx^i d\t  \\
& + \frac{2\ell}{r_c~\sqrt{V(r_c)}}\[ -\b_i dx^i dr+\(\tP+\frac{ r_c V'(r_c)}{2V(r_c)}\tP+ \frac{\b^2}{2}\)d\t dr+\hH(r)(r_c^2\tth_{\phi})d\t dr\]  \\
&+\frac{r^2}{r_c^2}\frac{V(r)}{V(r_c)}\(\frac{rV'(r)}{V(r)}-\frac{r_cV'(r_c)}{V(r_c)}\)\tP d\t^2
+ \frac{r^2}{r_c^2}  \(1-\frac{V\(r\)}{V\(r_c\)}\)\(\b^2d\t^2+ \b_i \b_j dx^i dx^j\)\\
& +\frac{r^2}{r_c^2} \frac{\hK(r)}{V(r_c)}(r_c^2\tth_{\phi})d\t dr
+ \frac{r^2}{r_c^2}\[{2r_c \hF(r) \sqrt{V(r_c)}} \si_{ij}+ r_c^2 \hF_{\phi}(r)(\si_{\phi})_{ij}\]dx^i dx^j,\\
\hA=&\frac{\ell U(r)}{r_c~\sqrt{V(r_c)}}\[\(1+\frac{\b^2}{2}\)+\(1-\frac{r U'(r)}{U(r)}+\frac{r_c V'(r_c)}{2V(r_c)}\)\tP\]d\t-\frac{\ell U(r)}{r_c~\sqrt{V(r_c)}}\b_i d\x^i\\
&+\A^{(ex)}_\t d\t+\A^{(ex)}_idx^i-\frac{\ell}{r_c}\frac{\m_0\hQ(r) }{\sqrt{V(r_c)}}\frac{r_h^{p-1}}{r^{p-1}} (r_c^2\tth_{\phi})d\t,\\
\tPh = &\phi(\x^a) +  r_c \sqrt{V(r_c)}\hF(r) \(\pc_\t\phi+\b^i\pc_i \phi\)  + r_c^2\hF_{\phi}(r)(\pc^i\pc_i\phi),
\end{aligned} \right.
\end{eqnarray}
and the shear components are given by
\begin{equation}\label{scaltwoderi}
\begin{split}
\si_{ij} =\pc_{(i} \b_{j)}- \frac{\d_{ij}}{p}\pc_k \b^k,\qquad
(\si_\phi)_{ij}=\pc_{(i} \phi\pc_{j)}\phi- \frac{\d_{ij}}{p} \tth_{\phi},\\
\end{split}
\end{equation}
where $\tth=\d^{ij}\pc_i
\b_j$,~$\tth_{\phi}=\d^{ij}\pc_i\phi\pc_j\phi$. At  order $\e^2$,
the temporal component of the forced NS equations in (\ref{tNS1})
turns out to be
\begin{equation}\label{FNS}
\begin{split}
\to_{_0}(r_c)\pc_i\b^i=0,\qq
\to_{_0}(r_c)={r_cV'(r_c)}/{(\ell\sqrt{V(r_c)})},
\end{split}
\end{equation}
which leads to the incompressible condition, and the spatial
component at order $\e^3$ is
\begin{equation}\label{FNS2} \begin{split}
\to_{_0}(r_c)\( \pc_\t \b_i + \b^j\pc_j \b_i\)-r_c\tp'_{_0}(r_c)\pc_i P -\eta_{_0}(r_c)\pc^2 \b_i=\f^{(\phi)}_i(r_c)+\f^{(q)}_i(r_c),
\end{split}
\end{equation}
where only the spatial components of external forces remain
\begin{align}
\f^{(\phi)}_{i}(r_c)\equiv&-\tch_{_0}(r_c)\(\b^j\pc_j\phi\pc_i\phi\)+\txi_{_0}(r_c)\(\pc^j\phi\pc_j\pc_i\phi\),\\
\f^{(q)}_{i}(r_c)\equiv&~ -\tnq_{_0}(r_c)\(\F^{(ex)}_{\t i}+\b^j\F^{(ex)}_{ji}\)-\tnq_{_0}(r_c)\(\F^{(bd)}_{\t i}+\b^j\F^{(bd)}_{ji}\).
\end{align}
If  further redefine
\begin{equation}
\tP_{[\o]}=-\frac{r_c\tp'_{_0}(r_c)}{\to_{_0}(r_c)}\tP,~~~
\tn_{[\o]}=\frac{\eta_{_0}(r_c)}{\to_{_0}(r_c)},~~~
{\f}^{(\phi)}_{[\o]i}=\frac{{f}^{(\phi)}_i(r_c)}{\to_{_0}(r_c)},~~~
{\f}^{(q)}_{[\o]i}=\frac{{f}^{(q)}_i(r_c)}{\to_{_0}(r_c)},
\end{equation}
we can obtain the incompressible forced NS equations as
\begin{equation}\label{tFNS1}
\begin{split}
 \pc_\t \b_i + \b^j\pc_j \b_i+\pc_i \tP_{[\o]} -\tn_{[\o]}\pc^2 \b_i=\f^{(\phi)}_{[\o]i}+\f^{(q)}_{[\o]i}
 ,\qq \pc_i \b^i=0.
\end{split}
\end{equation}
Consider the characteristic scale of perturbations $L\sim\e^{-1}$
and velocity $\b=\sqrt{\b_i\b^i}\sim\e$, the Reynolds number of the
fluid turns out to be
\begin{equation}
\begin{split}
\tmR_{e}(r_c)=\frac{\b L}{\tn_{_{0}}(r_c)}\propto\frac{1}{\tn_{_{0}}(r_c)}=\frac{\T_{_{0}}(r_c)}{\bar{D}_c(r_c)}=4\pi \T_{_{0}}(r_c)\[1+\frac{\tnq_{_{0}}(r_c)\m_{_{0}}(r_c)}{\T_{_{0}}(r_c) s_{_{0}}(r_c)}\].
\end{split}
\end{equation}
We can see that for uncharged black brane where the chemical
potential vanishes, the Reynolds number of the dual fluid is
proportional to the local temperature at the cutoff surface. Thus,
when the cutoff surface approaches the horizon, the local
temperature as well as the Reynolds number become larger and larger,
and the fluid may become unstable. This instability may relate to
the superluminal hydrodynamic sound modes when the cutoff surface is
sufficiently close to the horizon \cite{Brattan:2011my,
Marolf:2012dr}. In addition, as  in section \ref{sect:boundary2}, we
can  divide the electromagnetic forces into two parts:
 $\f^{(q)}_i(r_c)=\f^{(ex)}_i(r_c)+\f^{(bd)}_i(r_c)$, where
\begin{align}
\f^{(ex)}_i(r_c)
&=-\tnq_{_0}(r_c)\(\pc_\t \A^{(ex)}_i-\pc_i \A^{(ex)}_\t+ v^j \pc_j \A^{(ex)}_i-v^j \pc_i \A^{(ex)}_j\),\\
\f^{(bd)}_i(r_c)
&=\tnq_{_0}(r_c)\[\tm_{_0}(r_c)\(\p_\t \b_i+ \b^j \p_j \b_i\)-r_c\tm'_{_0}(r_c)\p_i P\].
\end{align}
Define
\begin{equation}
\begin{split}
\tP_{[s]}&=\frac{\tnq_{_0}(r_c)r_c\tm'_{_0}(r_c)-r_c\tp'_{_0}(r_c)}{\T_{_0}(r_c)s_{_0}(r_c)}\tP,
~~~{\f}^{(\phi)}_{[s]i}=\frac{{\f}^{(\phi)}_i(r_c)}{\T_{_0}(r_c)s_{_0}(r_c)},
~~~{\f}^{(ex)}_{[s]i}=\frac{{\f}^{(ex)}_i(r_c)}{\T_{_0}(r_c)s_{_0}(r_c)},
\end{split}
\end{equation}
and the dynamical shear viscosity
$\tn_{[s]}={\eta_{_0}(r_c)}/[{\T_{_0}(r_c)s_{_0}(r_c)}]$, the forced
NS equations (\ref{tFNS1}) then become
\begin{equation}\label{FNS1}
\begin{split}
 \pc_\t \b_i + \b^j\pc_j \b_i+\pc_i \tP_{[s]} -\tn_{[s]}\pc^2 \b_i=\f^{(\phi)}_{[s]i}+\f^{(ex)}_{[s]i}
 ,\qq \pc_i \b^i=0.
\end{split}
\end{equation}
 In this case, the external forces only come from the  additional electromagnetic field and dilaton field.
And the Reynolds number has the form
$\tmR_{e}(r_c)
\propto{\tn^{-1}_{[s]}(r_c)}=4\pi
\T_{_{0}}(r_c)$.

\section{Conclusions}
\label{sect:co} We have studied the thermodynamics and
non-relativistic hydrodynamics of the holographic fluid at a finite
cutoff surface. As a calculation example, we have considered the
case with the Gauss-Bonnet gravity. The isentropic flow and shear
viscosity of the dual fluid have been obtained. The radial Einstein
equation implies the isentropy of RG flow was first proposed in
\cite{Bredberg:2010ky}, and here we have generalized the discussion
to the case of the Gauss-Bonnet gravity. The isentropy of RG flow
can also be considered as an adiabatic process of the dual fluid.
Note that instead of the entropy associated with the holography
screen~\cite{Verlinde:2010hp,Cai:2010sz}, here the total entropy of
the dual fluid is the horizon entropy of the background black brane,
hence cutoff-independent. We have given a general formula
(\ref{shear}) on the ratio of shear viscosity over entropy density
of the fluid dual to the Gauss-Bonnet gravity. It shows that the
ratio is independent of the cutoff surface. Namely it does not run
with the cutoff surface.  This may explain why the membrane
paradigm~\cite{Parikh:1997ma,Jacobson:2011dz} and the AdS/CFT
correspondence~
\cite{Brigante:2007nu,Brigante:2008gz,Ge:2008ni,Cai:2008ph,Cai:2009zv,Ge:2009eh,Hu:2010sn,Hu:2011ze}
give the same result. This formula is also valid for the fluid dual
to the Rindler spacetime in Gauss-Bonnet gravity.

The main goal of this paper is to give the expressions of external force,
coming from the bulk matters, of holographic fluid in
the non-relativistic limit. To this end, we have considered an
Einstein-Maxwell-dilaton system with a negative cosmological
constant. By using the non-relativistic fluid expansion method, we
have solved the system up to the second order of non-relativistic
fluid expansion parameter $\e$, and obtained the incompressible
forced NS equations of the dual fluid at the AdS boundary and at a
finite cutoff surface, respectively. The concrete expressions of
external forces from the dilaton field and Maxwell field have been
given. Here we have taken an new scaling of the dilaton field $
\p_i\phi \sim \e^1$ in the non-relativistic limit, so that the
external force provided by the dilaton field $ f_i^{(\phi)} \sim
\e^3$, meeting  the scaling symmetry of the forced NS equations.
Actually, in the derivative expansion of the stress energy tensor,
these terms such as $\p_i\phi\p_j\phi$ are the second order
dissipative terms of the dual fluid \cite{Bhattacharyya:2008ji}, and
such terms may  appear in the superfluid
components~\cite{Sonner:2010yx}. However, in the non-relativistic
limit, we move them to the right-hand side of the NS equations as
external force terms. Note that here we have considered the case of
minimal coupling of dilaton field. It would be interesting to extend
this study to the case with non-minimal
coupling~\cite{Gouteraux:2011qh,He:2011hw,Cai:2012xh}.

It turns out the Reynolds number of dual fluid is proportional to
the local temperature of the cutoff surface in the uncharged case.
Thus when the cutoff surface approaches to the event horizon of the
black brane background, the local temperature and thus the Reynolds
number become larger and larger. This indicates that the dual fluid
will become unstable when the cutoff surface is close enough to the
event horizon. It would be interesting to further study the
stationary turbulence by using the forced NS equations derived in
this paper.
In \cite{Jejjala:2008jy,Fouxon:2009rd,Eling:2010vr,Jejjala:2010tq,Evslin:2010ij},
the problems of turbulence have been studied via a holographic
description with gravity.
It was also shown that the nonlinear evolution of anti-de
Sitter space might be unstable, and the energy of
perturbations would be transformed to smaller and smaller scales
like the turbulence energy cascades \cite{Bizon:2011gg,Dias:2011ss}.
Thus it would be interesting to establish the holographic turbulence using
the AdS/CFT correspondence \cite{Maldacena:1997re,Aharony:1999ti}.

\section*{Acknowledgements}
This work was supported in part by the National Natural Science
Foundation of China (No.10821504, No.10975168 and No.11035008), and
in part by the Ministry of Science and Technology of China under
Grant No. 2010CB833004. We would like to thank Jarah Evslin,
Xian-Hui Ge, Song He, Bin Hu, Ya-Peng Hu, Yi Ling, Da-Wei Pang,
Yong-Hui Qi, Yu Tian, Xiao-Ning Wu, Hai-Qing Zhang  and Yang Zhou
for illuminating conversations during this work. Z.-Y. Nie would like to
thank  the Conference on Cold Materials, Hot Nuclei and Black Holes:
Applied Gauge/Gravity Duality, held at the Abdus Salam International
Centre for Theoretical Physics, for their hospitality.

\end{document}